\documentclass[conference]{IEEEtran}

\makeatletter
\newcommand\semihuge{\@setfontsize\semihuge{22.3}{22}}
\makeatother

\usepackage{algpseudocode}
\usepackage{algorithm}
\usepackage{algorithmicx}

\usepackage{lipsum} % For algorithm smape after or before:
\usepackage{amsfonts}

\usepackage{multirow}
\usepackage[dvips]{color}
\usepackage{comment}
\usepackage{todonotes}
\usepackage{epsf}
\usepackage{epsfig}
\usepackage{times}
\usepackage{epsfig}
\usepackage{graphicx}
\usepackage{mathtools}
\usepackage{mathrsfs}
\usepackage{amssymb}
\usepackage{pdfpages}
\usepackage{epstopdf}
\usepackage{float}
\newfloat{algorithm}{t}{lop}
\usepackage[algo2e]{algorithm2e}
\usepackage{subfig}

\usepackage{dsfont}
\usepackage[multiple]{footmisc}
\usepackage{lettrine} 
\usepackage{amsmath,epsfig,amssymb,algorithm,algpseudocode,amsthm,cite,url}

\usepackage{caption}

\allowdisplaybreaks
\usepackage{csquotes}
\usepackage{multirow}

\DeclareMathOperator*{\argmin}{arg\,min}

\usepackage{verbatim}
\usepackage{subfig}
\usepackage[english]{babel}
\usepackage{amsmath,amssymb}

\usepackage{verbatim}

\newtheorem{corollary}{Corollary}
\newtheorem{theorem}{\bf Theorem}

\newtheorem{lemma}{\bf Lemma}

\IEEEoverridecommandlockouts	
\begin{document}
	\bstctlcite{IEEEexample:BSTcontrol}
\title{\semihuge  Federated Learning on the Road Autonomous Controller Design for Connected and Autonomous Vehicles }

\author{\IEEEauthorblockN{ Tengchan Zeng\IEEEauthorrefmark{1},~\IEEEmembership{Student Member,~IEEE,}, Omid Semiari\IEEEauthorrefmark{2},~\IEEEmembership{Member,~IEEE,} \\Mingzhe Chen \IEEEauthorrefmark{3},~\IEEEmembership{Member,~IEEE,} Walid Saad\IEEEauthorrefmark{1},~\IEEEmembership{Fellow,~IEEE,} and Mehdi Bennis\IEEEauthorrefmark{4},~\IEEEmembership{Fellow,~IEEE,}}
	\IEEEauthorblockA{
		\small \IEEEauthorrefmark{1} Wireless@VT, Department of Electrical and Computer Engineering, Virginia Tech, Blacksburg, VA, USA,\\ Emails:\url{{tengchan , walids}@vt.edu}.\\
		\IEEEauthorrefmark{2} Department of Electrical and Computer Engineering, University of Colorado, Colorado Springs, CO, USA, \\Email: \url{osemiari@uccs.edu}.\\
		\IEEEauthorrefmark{3} Department of Electrical Engineering, Princeton University, Princeton, NJ, USA, Shenzhen Research Institute of Big Data (SRIBD) \\and the Future Network of Intelligence Institute (FNii), Chinese University of Hong Kong, Shenzhen, China, \\Email: \url{mingzhec@princeton.edu}.\\
		\IEEEauthorrefmark{4} Centre for Wireless Communications, University of Oulu, Oulu, Finland, Email:
		\url{mehdi.bennis@oulu.fi}.
		\thanks{\textcolor{black}{A preliminary version of this work appears in the proceeding of IEEE Conference on Decision and Control (CDC), 2021 \cite{icc2021Zeng}. 
			}}}
}
\maketitle

\begin{abstract}
	The deployment of future intelligent transportation systems is contingent upon seamless and reliable operation of connected and autonomous vehicles (CAVs).
	One key challenge in developing CAVs is the design of an autonomous controller that can accurately execute near real-time control decisions, such as a quick acceleration when merging to a highway and frequent speed changes in a stop-and-go traffic.   
	However, the use of conventional feedback controllers or traditional learning-based controllers, solely trained by each CAV's local data, cannot guarantee a robust controller performance over a wide range of road conditions and traffic dynamics.
	In this paper, a new federated learning (FL) framework enabled by large-scale wireless connectivity is proposed for designing the autonomous controller of CAVs. 
	In this framework, the learning models used by the controllers are collaboratively trained among a group of CAVs. 
	To capture the varying CAV participation in the FL training process and the diverse local data quality among CAVs, a novel dynamic federated proximal (DFP) algorithm is proposed that accounts for the mobility of CAVs, the wireless fading channels, as well as the unbalanced and non-independent and identically distributed data across CAVs.
	A rigorous convergence analysis is performed for the proposed algorithm to identify  how fast the CAVs converge to using the optimal autonomous controller.
	In particular, the impacts of varying CAV participation in the FL process and diverse CAV data quality on the convergence of the proposed DFP algorithm are explicitly analyzed.  
	Leveraging this analysis, an incentive mechanism based on contract theory is designed to improve the FL convergence speed.
	Simulation results using real vehicular data traces show that the proposed DFP-based controller can accurately track the target CAV speed over time and under different traffic scenarios. 
	Moreover, the results show that the proposed DFP algorithm has a much faster convergence compared to popular FL algorithms such as federated averaging (FedAvg) and federated proximal (FedProx).
	The results also validate the feasibility of the contract-theoretic incentive mechanism and show that the proposed mechanism can improve the convergence speed of the DFP algorithm by $40$\% compared to the baselines.

\end{abstract}

\section{Introduction}
As a key component of tomorrow's intelligent transportation systems (ITSs), connected and autonomous vehicles (CAVs) are emerging as a promising solution to reduce traffic accidents, alleviate road congestion, and increase transportation efficiency.
CAVs leverage sensors together with wireless systems to increase their situational awareness and improve their motion planning and automatic control. 
However, to operate full-fledged CAVs, we need to address a number of challenges, ranging from providing seamless wireless connectivity to designing reliable controllers.
Among these challenges, designing an autonomous controller to achieve target movements for CAVs is critical in order to allow a CAV to accomplish its target tasks and operate safely.
In particular, a CAV's controller must accurately execute navigation decisions so that the CAV can quickly adapt to the dynamic road traffic \cite{7490340}.
For example, the controllers must generate frequent slow-down and speed-up for CAVs in a stop-and-go traffic, whereas a rapid acceleration will be the target output for the controllers when CAVs merge into highways.  

\subsection{Motivation and Related Works}

There are two common methods to design an autonomous controller for CAVs. 
The first method uses a conventional feedback controller. 
In particular, the conventional feedback controller first determines the CAV's dynamic models (e.g., the tire model \cite{7225830}) as well as the road conditions (e.g., road slope \cite{5721826} and slip ratio between the road and tire \cite{nam2015wheel}), and then optimizes the controller design based on these settings. 
However, due to various types of roads, dynamic road traffic, and varying weather, the road conditions will change constantly.
Hence, a conventional feedback controller cannot guarantee the controller performance over a wide range of environmental parameter changes.
To ensure that the CAVs can adapt to changing road conditions, the second method relies on the use of adaptive controllers, based on machine learning (ML), for the CAV's autonomy.
For example, in \cite{doi:10.1146/annurev-control-090419-075625}, the authors propose a learning-based model predictive control (MPC) design where the recorded trajectory data is trained to optimize the parameterization of the MPC controller that leads to the optimal closed-loop performance.
The authors in \cite{nie2018longitudinal} use a radial basis function neural network to design an adaptive proportional integral derivative (PID) controller so as to achieve accurate longitudinal movement for autonomous vehicles.
In \cite{han2017lateral}, a multilayer perceptron is used to design an adaptive lateral controller for improving path tracking in autonomous vehicular systems.
However, when using ML methods for adaptive controller design, the local data can be insufficient to train the learning model due to the limited on-chip memory available on board CAVs \cite{lin2018architectural}. 
In fact, because of the limited storage, an individual CAV can only store data pertaining to its most recent travels, and this data can be easily skewed and of poor quality.
Hence, when changing to a new traffic environment or when a CAV encounters less frequently occurring events (e.g., traffic accidents), a controller solely trained by the local data can fail to adapt to such dynamics.
An effective controller design will thereby hinge on training the ML model using the data collected by more than one CAV.
In other words, a cooperative learning framework among multiple CAVs will be needed for properly designing the autonomous controller of a CAV.

To this end, one can leverage the wireless connectivity in CAVs and use federated learning (FL) to enable a network of CAVs to collaboratively train the learning models used by their controllers \cite{9169921}.
In FL, the CAVs can train the controller models based on their local data available at their local memory and, then, a parameter server, such as a base station (BS), can aggregate the trained controller models from CAVs. 
These processes will be repeated among the CAVs and parameter server iteratively until all controllers converge to the optimal learning model.
In this way, the learning model can be collaboratively trained among multiple CAVs, and \emph{such a trained model can enable a particular CAV's controller to adapt to new traffic scenarios unknown to the CAV but already experienced by other CAVs in the past}. 
For example, as shown in Fig. \ref{system_model}\subref{system_model1}, the CAVs participating in an FL process can learn from each other to operate in a wide range of scenarios, such as accident, traffic jam, and roadwork areas.
Moreover, the FL process is naturally privacy-preserving as the CAVs do not share their local data, e.g., the history trajectory.  

To reap all these benefits, we need to address a number of challenges.   
First, due to the CAV's mobility and uncertainty of wireless channels, the participation of CAVs in the FL process will vary over time, and hence, it can be challenging to guarantee a good training performance.
Second, because of the unbalanced and non-independent and identically distributed (non-IID) local data across CAVs, the data quality among CAVs will be different and such diverse data quality can impact the FL convergence. 
%An effective FL framework for the CAV's controller design must solve these two challenges.
Third, when implementing FL for autonomous controller design, it is necessary to design an effective mechanism that incentivizes the CAVs to participate the ML training. 
In particular, the designed incentive mechanism must offer a reward to CAVs so as to compensate the cost of the energy spent on the local training and uplink transmission.
Meanwhile, considering the diverse local data quality at the CAVs, the incentive mechanism must be designed in a way to motivate the participation of CAVs with good data quality and prevent CAVs with poor data quality from engaging in the FL process. 
Such incentive mechanism design becomes more challenging in the context of FL since there exists an \emph{information asymmetry} between the parameter server and CAVs. 
That is, only the CAVs can know their own data quality while the parameter server cannot access to the CAVs' local data. 

To design an effective incentive mechanism in FL, there are a number of works using game-theoretic and learning concepts (such as deep reinforcement learning \cite{8963610} and Stackelberg game \cite{9247530}) where the parameter server offers rewards to the local users for their participation in the FL process. 
In particular, in each round, the parameter server will communicate iteratively with the local users to determine the payment plan which ensures a target number of users participating in the FL process while minimizing the total payment at the parameter server.
However, this process can be time-consuming and could result in a non-negligible delay, posing a safety threat to the real-time operation of CAVs. 
Meanwhile, there are works that use the framework of contract theory to design realistic incentive strategies, as done in \cite{8964354,9057543,8851649}.
These works generally group the users into different types and design a contract for each user type. 
Users will then self-reveal their types by choosing the contracts especially designed for their type.
Nevertheless, these prior works group users according to simple definitions of data quality (e.g., image quality \cite{8964354} and accuracy \cite{9057543} and \cite{8851649}). 
Such metrics can only capture 
the data quality at the level of individual data samples while ignoring the data size and distribution.
In addition, the works in \cite{8964354,9057543,8851649} assume that any CAV participation in the FL process will improve the overall convergence. 
In fact, due to the diverse data quality, the convergence of the FL process can be impeded by CAVs that have poor quality data.
Hence, for CAVs, one must design an incentive mechanism that can offer rewards for the energy cost at CAVs while also accelerating the convergence of the controller design.

\subsection{Contributions and Outcomes}
The \textit{main contribution} of this paper is a novel FL framework that enables CAVs to collaboratively learn and optimize their autonomous controller design in presence of wireless link uncertainties and environmental dynamics.
In particular, we propose a dynamic federated proximal (DFP) algorithm. 
Different from the conventional FL-based methods in \cite{konevcny2016federated,9210812,kim2022,smith2017federated}, we consider both the time-varying participation of CAVs in the FL process and non-IID local data across CAVs when performing FL in a vehicular communication network.

To speed up the convergence of the proposed DFP algorithm, we design a contract-theoretic incentive mechanism to allocate the transmit power for CAVs.
In particular, we model the interactions between the parameter server and the CAVs as a labor market where the parameter server is the employer and CAVs are the employees. 
Here, we first mathematically capture the data quality for each CAV according to how the local data affects the overall convergence. 
Then, according to the data quality, we partition the CAVs into different types, and design a contract, i.e., transmit power-reward bundle, for each CAV type. 
These contracts are designed in a way to improve convergence by motivating CAVs with good data quality to allocate more transmit power for the uplink transmission in the FL process while CAVs with poor data quality will spend less or no transmit power.
Using real vehicular data traces, i.e., the Berkeley deep drive (BDD) data \cite{yu2020bdd100k} and the dataset of annotated car trajectories (DACT) \cite{10.1145/3152178.3152184}, we show that the controller trained by our proposed algorithm can track the target speed over time and under different traffic scenarios (e.g., traffic accidents, traffic congestion, and roadwork zones).  
Also, when using the proposed algorithm for the controller design, the distance error is shown to be two times smaller than controllers solely trained by the local data.
In addition, simulation results show that, the proposed algorithm can achieve a faster convergence than FedAvg and federated proximal (FedProx) algorithms, leading to a quick adaptation to the traffic dynamics. 
Furthermore, the results validate the feasibility of the proposed contract-theoretic incentive mechanism and show that the mechanism can improve the convergence speed of DFP algorithm by 40\% compared with the baseline schemes.
\emph{To the best of our knowledge, this is the first work that develops an FL framework to optimize the autonomous controller design for CAVs.}

The rest of the paper is organized as follows. Section \uppercase\expandafter{\romannumeral2} presents the control, learning, and communication models. 
The proposed algorithm and its convergence proof are studied in Section \uppercase\expandafter{\romannumeral3}. 
In Section \uppercase\expandafter{\romannumeral4}, the contract-theory based incentive mechanism is introduced. 
Section \uppercase\expandafter{\romannumeral5} provides the simulation results and conclusions are drawn in Section \uppercase\expandafter{\romannumeral6}.

\section{System model}
\begin{figure}[!t]
	\centering 
	%\vspace{-0.05in}
	\subfloat[]{%
		\includegraphics[width=2.3in,height=1.8in]{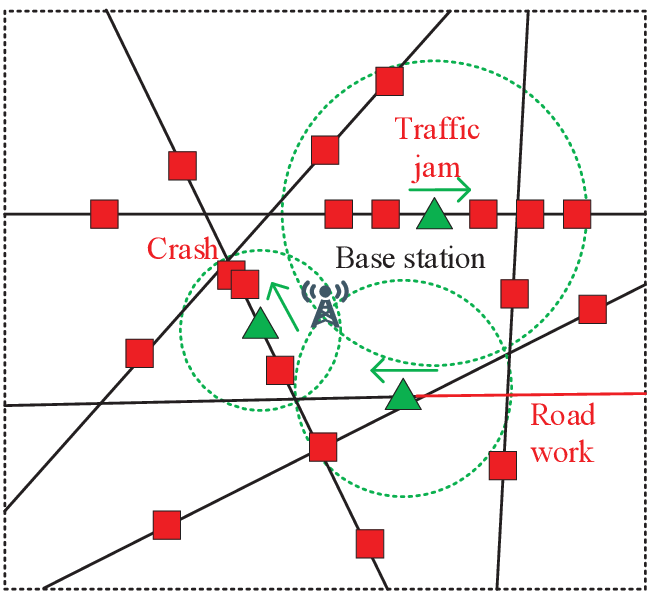}\label{system_model1}
	}	

	\subfloat[]{%
		\includegraphics[width=2.3in,height=1.85in]{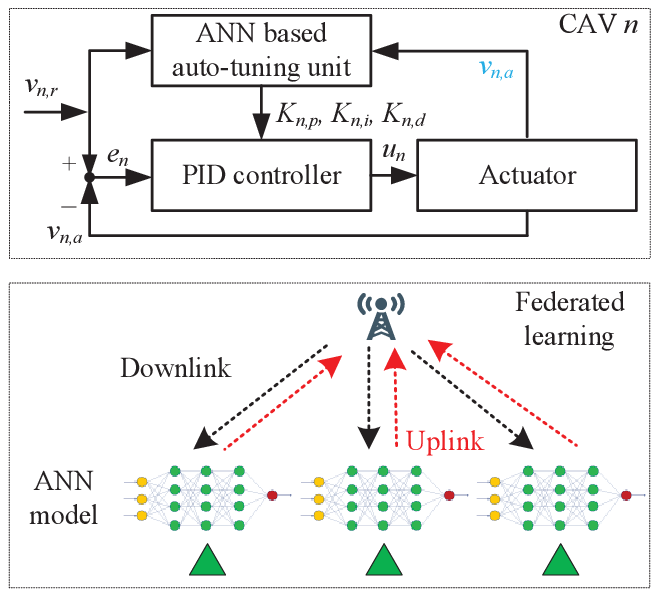}\label{system_model2}
	}	
	\caption{Illustration of our system model. The traffic model is presented in (a) where green triangles and red squares, respectively, represent CAVs that do and do not participate in the FL process. The adaptive controller and learning models are shown in (b).}
	\label{system_model}	
	\vspace{-0.1in}
\end{figure}

Consider a cellular BS serving a set $\mathcal{N}$ of $N$ CAVs with the same type of vehicle dynamics that move along a road system, as shown in Fig. \ref{system_model}\subref{system_model1}. 
Each CAV will perceive its surrounding environment and accordingly adjust the controller decisions in order to achieve the target movement.
FL is used to learn the controller so that CAVs can automatically change their control parameters, execute control decisions, and adapt to their local traffic. Note that, to adapt the road traffic at different times, the frequency that FL is used to update the controller design will be time-varying.
We will next introduce the controller, communication, and learning models used for our FL-based autonomous controller design framework. 

\subsection{Adaptive Longitudinal Controller Model}
To perceive their surrounding environment, CAVs will use sensors and communicate with nearby CAVs and BS. 
This environmental perception enables the longitudinal controller of each CAV to automatically adjust its acceleration or deceleration and maintain a safe spacing and target speed.
Due to the simplicity and ease of implementation of a PID controller, we assume that it is used by CAVs to control their longitudinal movement. Then, the acceleration $u_{n}(t)$ of vehicle $n \in \mathcal{N}$ at time sample $t$ is \cite{dos2009improved}
\begin{align}
	\label{acceleration} 
	u_{n}(t)\!=& u_{n}(t\!-\!1)\!+\!\! \left(\!\!K_{n,p}\!+\!K_{n,i} \Delta t \!+\! \frac{K_{n,d}}{\Delta t}\!\!\right)e_n(t)\!+\\&\left(\!\!-K_{n,p}\!-\!\frac{2K_{n,d}}{\Delta t}\!\!\right)e_n(t\!-\!1) \!+\!\!\frac{K_{n,d}}{\Delta t} e_n(t\!-\!2),
\end{align}
where non-negative coefficients $K_{n,p}, K_{n,i},$ and $K_{n,d}$ are, respectively, the proportional gain, integral time constant, and derivative time constant used by the PID controller at CAV $n\in \mathcal{N}$. $\Delta t$ is the sampling period and  $e_n(t)=v_{n,r}(t)-v_{n,a}(t)$ captures the difference between the target reference speed $v_{n,r}(t)$ and the actual speed $v_{n,a}(t)$ at sample $t$. 
Note that the target reference speed is decided by the motion planner in the CAV based on the environmental perception\footnote{As the motion planner design has been extensively studied by the prior art and is not the main scope of this work, we omit details about the process of choosing the target speed and we refer readers to \cite{7490340} for further details.}.

According to (\ref{acceleration}), we can calculate the actual speed at sample $t+1$ as $v_{n,a}(t+1)=v_{n,a}(t)+ u_{n}(t)\Delta t$ and the distance traversed between samples $t$ and $t+1$ as $d_{n,p} = \frac{v_{n,a}(t+1)+v_{n,a}(t)}{2}\Delta t$. 
Clearly, achieving the target speed and safe spacing will depend on the control parameter setting of the PID controller. 
Hence, it is imperative to adjust these control parameters adaptively to deal with varying traffic dynamics and road conditions. 
To this end, as shown in Fig. \ref{system_model}\subref{system_model2}, we assume that the CAV will use an adaptive PID controller enabled by an artificial neural network (ANN) based auto-tuning unit.
Here, we use an ANN because it is capable of capturing the nonlinear relationship between the PID control parameter setting and the longitudinal controller performance (i.e., velocity errors) \cite{doi:10.1146/annurev-control-090419-075625}. 
Hence, the ANN-based auto-tuning unit can dynamically adjust the control parameters, and the CAVs can adapt to varying traffic scenarios.
Meanwhile, to guarantee that the PID control parameters will be always positive, we use the sigmoid function as an activation function in the ANN.
In this case, to adapt to various traffic conditions, the CAV will train the auto-tuning unit by using the back-propagation algorithm over its own local data and adjust the control parameters accordingly.
This is an emerging approach for adaptive controller design, as discussed in \cite{doi:10.1146/annurev-control-090419-075625,nie2018longitudinal,han2017lateral}.
\subsection{FL Model}
The ANN based auto-tuning unit in Fig. \ref{system_model}\subref{system_model2} can adaptively tune the PID control parameters to achieve the target speed. 
However, the CAV's local training data (e.g., camera data containing the longitudinal movement) is constrained by the onboard memory of the CAV, and, thus, the information that can be stored will be limited to a few traffic scenarios. 
For example, for CAVs driving on the highway, the longitudinal movement data captured by the camera will be mostly high speed data. 
As a result, the trained controller can only operate in the highway scenario and cannot adapt to stop-and-go traffic with frequent stops and accelerations when CAVs exit the highway and drive in urban settings.
In other words, by solely training the local data for the auto-tuning unit, the controller can only work in limited traffic scenarios but not in presence of a more general traffic pattern which could jeopardize the safe operation of CAVs.
To address this challenge, we can use the wireless connectivity of CAVs to build a cooperative, learning-based training framework, i.e., FL, among multiple CAVs for the controller design. 

Here, we consider that CAVs will engage in an FL process to collaboratively learn the ANN auto-tuning units for their adaptive controller design. 
In particular, a wireless BS, operating as a parameter server, will first generate an initial global ANN model parameter $\boldsymbol{w}_{0}$ for the auto-tuning unit and send it to all CAVs over a downlink broadcast channel.
Then, in the first communication round, CAVs will use the received model parameters $\boldsymbol{w}_{0}$ to independently train their own model based on their local data for $I$ iterations.
Note that, due to the temporal correlation within the road traffic, the CAV will train the ANN with initial training parameters (i.e., weights and bias) that are close to the target values, guaranteeing stability \cite{ge2013stable}.
Meanwhile, the motion planner can take into account the stability of the controller when designing target velocity traces \cite{7353382}, thereby further enhancing the controller's stability.
In the uplink, the CAVs transmit their trained model parameters to the BS. 
Next, the BS will aggregate all the received local model parameters to update the global model parameters which are then sent back to all CAVs over the downlink broadcast channel. 
This FL process is repeated over uplink-downlink channels and the local and global ANN models are sequentially updated in the following communication rounds. 
Ultimately, the ANN model parameters used by the CAVs will converge to the optimal model after solving the following optimization problem that captures the FL process \cite{9210812}: 
\begin{align}
\label{optimization}
&\argmin_{\boldsymbol{w}^{(1)},...,\boldsymbol{w}^{(N)}\in \mathbb{R}^d}\sum_{n=1}^{N}\sum_{i=1}^{s_{n}}\frac{s_n}{s_N} f_{n}(\boldsymbol{w}^{(n)},\xi_{i}), \\ 
&\hspace{0.22in}\text{s.t.} \hspace{0.1in} \boldsymbol{w}^{(1)} = \boldsymbol{w}^{(2)} = . . . = \boldsymbol{w}^{(N)} =\boldsymbol{w}, \vspace{-0.1cm}
\end{align}
where $s_N= \sum_{n\in \mathcal{N}}s_{n}$ is the size of all training data available at the local memory of CAVs with $s_{n}$ being the size of the local data at CAV $n$. $f_n(\boldsymbol{w}^{(n)},\xi_{i})$ is the loss function of CAV $n$ when using the ANN model parameters $\boldsymbol{w}^{(n)}$ in the auto-tuning unit for the selected data $\xi_{i}$.
Note that, the loss function plays a pivotal role in determining the performance of the trained auto-tuning unit. 
The loss function used for the controller design can be either convex \cite{4469948} or non-convex \cite{liu2014solving}. 
We assume $f(\boldsymbol{w})$ to be the value of the objective function in (\ref{optimization}) when $\boldsymbol{w}^{(n)}=\boldsymbol{w}, n \in \mathcal{N}$.

When training the local ANN models at CAVs, we can calculate the energy consumption for CAV $n \in \mathcal{N}$ in each communication round as $E_{n,\text{comp}}= \kappa c \phi^{2} \bar{s} I$, where $\kappa$ is the energy consumption coefficient that depends on the computing system and $\bar{s}$ is the size of training data at the local iteration. $c$ is the number of computing cycles needed per bit, and $\phi$ is the frequency of the central processing unit (CPU) clock of CAV.
Accordingly, we can obtain the computing delay as $t_{n,\text{comp}}=I\frac{\bar{s}c}{\phi}$. 
Due to the mobility of CAVs and the wireless fading channels, some CAVs cannot finish their local training and uplink transmission within the duration $\bar{t}$ of the communication round.
With this in mind, next, we present the communication model used to determine whether the locally trained model at a particular CAV can be used in the model aggregation or not.  %\vspace{-0.07in}
\subsection{Communication Model}
For the uplink transmissions, we consider an orthogonal frequency-division multiple access (OFDMA) scheme where each CAV in set $\mathcal{N}$ will use a unique orthogonal resource block to transmit the trained ANN model parameters to the BS. 
In particular, the BS will allocate orthogonal subcarriers to CAVs so as to avoid interference between concurrent uplink transmissions. 
This is a practical assumption given that a single BS will only service a handful of CAVs that are of the same vehicular type.
The data rate for the link between a CAV $n \in \mathcal{N}$ and the BS will be
\begin{align}
\label{uplink}
r_{n} = B \log_{2} \left(1+\frac{P_n h_{n} d_{n}^{-\alpha}}{\delta_{n} + B N_0}\right),
\end{align}
where $B$ is the bandwidth of each resource block, $P_{n}$ is the transmit power of CAV $n$, and $h_{n}$ is the wireless fading channel gain.
In particular, since line-of-sight links between CAVs and the BS do not always exist, we model these channels as independent Rayleigh fading channels \cite{acosta2006doubly}.
Moreover, $d_{n}$ is the distance between CAV $n$ and the BS, $\alpha$ is the path-loss exponent, and $N_0$ is the noise power spectral density. 
In addition, $\delta_{n}=\sum_{j \not\in \mathcal{N}}  P_j h_{j}d_{j}^{-\alpha}$ is the received interference power generated by CAVs in other cells that share the same resource block with CAV $n$. 
%It is clear that both $h_{n}$ and $d_{n}$ will change over time due to the time-varying fading channels and the CAV's acceleration or deceleration according to (\ref{acceleration}). 
From (\ref{uplink}), the uplink transmission delay for CAV $n\in\mathcal{N}$ can be calculated as 
$t_{n,\text{comm}} = \frac{s(\boldsymbol{w}^{(n)})}{r_{n}}$, 
where $s(\boldsymbol{w}^{(n)})$ is the size of the data packet that depends on the trained model parameters, $\boldsymbol{w}^{(n)}$, transmitted by CAV $n$. 
The uplink energy consumption is $E_{n,\text{comm}}=P_{n}\hat{t}$ where $\hat{t}$ is calculated by the product of the total number of data symbols and the symbol duration.

In the downlink, since the BS can have a higher transmit power and a larger bandwidth, the downlink transmission delay is considered to be negligible compared to the uplink transmission delay, as assumed in \cite{8737464}. 
In addition, given the higher computing power of BSs, the computing delay at the BS can be ignored.
Hence, to identify whether the local learning model update from CAV $n \in \mathcal{N}$ can be used for the model aggregation in the BS,  we can compare the time for uplink transmission and local computing at the CAV with the duration $\bar{t}$ of the communication round. 
In this case, the probability that CAV $n \in \mathcal{N}$ participates at communication round $t$ of FL (i.e., the locally trained model at CAV $n$ is used in the model aggregation) will be given by $p_{n,t}=\mathbb{P}(t_{n,\text{comp}}+t_{n,\text{comm}}\leq \bar{t})$.

When developing the FL framework for the CAV's controller design, we need to address a number of challenges. 
The first challenge is that the BS can only aggregate a varying subset of CAVs to update the global model at each communication round as a result of the mobility of the CAVs and the uncertainty of wireless channels.
A fast convergence for the controller design will be challenging to achieve when the participation of the CAVs in the FL process varies over time \cite{mingzhe}. 
Meanwhile, as the local data is generated under various traffic scenarios and road incidents, its distribution and size will be different across CAVs.  
Hence, the second challenge will be mitigating the impact of the non-IID and unbalanced local data on the convergence of the controller design.
In the following section, we will propose a novel FL algorithm to tackle these two challenges.

Moreover, due to the energy cost in the model training and uplink transmission, another challenge will be designing an incentive mechanism that encourages CAVs to participate in the proposed FL algorithm. 
However, to improve the convergence performance, the incentive mechanism should only  motivate a subset of CAVs which can improve the convergence process of controller design, while preventing other CAVs that impede the convergence from engaging in the FL process.
Such an incentive mechanism is of great importance for enabling CAVs to quickly adapt to the local traffic dynamics when exploiting our proposed FL algorithm. 
Next, to address this challenge,  we will use the insights obtained from the convergence study of the proposed FL algorithm and design a contract-theoretic incentive mechanism.

\section{Dynamic Federated Proximal Algorithm for CAV Controller Design} 
To address the challenges imposed by the varying CAVs' participation in the learning process and the non-IID and unbalanced data, we propose a new DFP algorithm.
In particular, we study how the mobility of the CAVs, wireless fading channels, and the diverse local data affect the convergence of the learning model. 
Here, we will first introduce the proposed DFP algorithm and then study its convergence. 
\subsection{Proposed Dynamic Federated Proximal Algorithm}
The proposed algorithm is summarized in Algorithm \ref{Alg1}. 
In particular, we assume that the CAVs will run $I$ iterations of stochastic gradient descent (SGD) at each round. 
In each iteration of SGD, CAV $n\in\mathcal{N}$ will solve the following optimization problem that minimizes the sum of the loss of a randomly selected local training sample $\xi \in \mathcal{S}_{n}$ and an $L_{2}$ regularizer: $
\argmin_{\boldsymbol{w}\in \mathbb{R}^n} f_{n}(\boldsymbol{w},\xi) + \frac{\gamma_t}{2} ||\boldsymbol{w}-\boldsymbol{w}_t||^2, \xi \in \mathcal{S}_{n}, 
$
where $\gamma_t$ is the coefficient of the regularizer and $\boldsymbol{w}_t$ captures the received learning model parameters from the BS at communication round $t$.
Different from FedAvg algorithm \cite{konevcny2016federated}, we introduce the $L_2$ regularizer to guarantee that the trained model parameters $\boldsymbol{w}$ of CAV $n\in \mathcal{N}$ will be close to $\boldsymbol{w}_{t}$ during the local training, reducing the variance introduced by the non-IID and unbalanced data.
Meanwhile, in contrast to popular FL algorithms, such as FedProx \cite{li2020federated}, we explicitly consider the impact of CAVs' mobility and uncertainty of wireless channels and model the participation probability as a dynamic variable for each CAV at each communication round.
After $I$ iterations of SGD at communication round $t$, we obtain the trained model parameters of CAV $n$ as follows:
\begin{align}
	\label{6}
	f_{n}\!\left(\!\boldsymbol{w}_{t+1,I}^{(n)}\!\right) \!=\! \boldsymbol{w}_{t} \!+\! \eta_{t}\!\! \sum_{i=0}^{I\!-\!1}\! \left(\!\nabla f_{n}\!(\boldsymbol{w}_{t,i}^{(n)},\xi_{i})\!+\!\gamma_{t}(\boldsymbol{w}_{t,i}^{(n)}\!-\!\boldsymbol{w}_{t})\!\right), 
\end{align}
where $\boldsymbol{w}_{t,0}^{n}=\boldsymbol{w}_{t}, n\in\mathcal{N}$.

\subsection{Convergence of the Proposed DFP Algorithm}
Next, we perform a convergence study to determine how fast CAVs converge to using the optimal model in (\ref{optimization}) when exploiting the DFP algorithm. 
Unlike the convergence study done by existing works such as  \cite{konevcny2016federated} and \cite{li2020federated}, we need to consider how both the dynamic participation probability of CAVs and the $L_{2}$ regularizer in the local training affect the convergence.
To this end, we make the following standard assumptions:
\begin{itemize}
	\item The gradient $\nabla  f_{n}(\boldsymbol{w}), n \in \mathcal{N}$, is uniformly Lipschitz continuous in terms of $\boldsymbol{w}$ with positive parameter $L$. 
	\item The upper bound of the variance of SGD with respect to the full gradient descent of each CAV $n\in \mathcal{N}$ is $
	\mathbb{E}_{\xi \in \mathcal{S}_{n}}||\nabla f_{n}(\boldsymbol{w},\xi) - \nabla f_{n}(\boldsymbol{w})||_{2}^{2} \leq \sigma^2, \forall n\in \mathcal{N}, \forall \boldsymbol{w}\in \mathbb{R}^{d}$, 
	where $\sigma^2$ is the upper bound.
\end{itemize}
Both assumptions are commonly used in the convergence study of machine learning algorithms (e.g., see \cite{doi:10.1137/16M1080173}).
The first constraint can be satisfied by some popular loss functions used in control theory, such as the squared error loss function.
The second constraint is often adopted in stochastic optimization where the gradient estimator is always assumed to have a bounded variance. 
In the autonomous controller design problem, the second constraint can be justified by the fact that CAVs have limited acceleration and deceleration capabilities.
Using these two assumptions, we can bound the expected loss function at communication round $t+1$ as shown by the following theorem. 

\setlength{\textfloatsep}{4pt}
\begin{algorithm}[!t]
	
	\textbf{Iutput:} {$\mathcal{N}$, $\mathcal{N}_{t}$, $\mathcal{S}_n$, $\eta_{t}$, $\boldsymbol{w}_{0}$, $I$, $u_t$, $\gamma_t$, $s_n, n = 1, ..., N$}
	
	\textbf{Output:} { ANN-based auto-tuning unit $\boldsymbol{w}$ for the CAV's controller}
	
	{\For{$t = 0,...,T-1$}{
			
			1. The BS sends $\boldsymbol{w}_t$ to all CAVs over broadcast downlink channels.
			
			2. CAV $n\in \mathcal{N}$ updates $\boldsymbol{w}_t$ for $I$ iterations of SGD with a step size as $\eta_{t}$ in (\ref{6}) and obtain $\boldsymbol{w}_{t+1,I}^{(n)}$ which will be sent to the BS.
			
			3. Due to the mobility and wireless fading channels, the BS can only aggregate the trained model parameters from a subset $\mathcal{N}_{t}$ of $N_t$ CAVs and update the global model parameters as 
			$\boldsymbol{w}_{t+1}=\sum_{n\in \mathcal{N}_t} \frac{s_n}{s_{N_t}}\boldsymbol{w}^{(n)}_{t+1,I}$ with $s_{N_t}=\sum_{n\in \mathcal{N}_t}s_n$. 
	}}
	\caption{Dynamic Federated Proximal (DFP) Algorithm}
	\label{Alg1}
\end{algorithm}

\begin{theorem}
	\label{theorem1}
	%\vspace{-0.03in}
	\emph{Given that the BS sends the global learning model parameters $\boldsymbol{w}_{t}$ to all CAVs at communication round $t$, an upper bound for the expected loss function at communication round $t+1$ can be written as\vspace{-0.1in}}
	\begin{align}
		\label{major1}
		&\mathbb{E}_{\xi, n}(\!f(\boldsymbol{w}_{t\!+\!1})\!)\!\leq\!\! f(\boldsymbol{w}_{t})  \!-\!\!\frac{(\eta_{t}\!\!+\!\!\gamma_{t}\eta_{t})\!\sum_{n=1}^{N}p_{n,t}s_n^2I||\nabla f_{n}(\boldsymbol{w}_{t})||_2^{2}}{2s_{N}\sum_{j=1}^{N}p_{j,t}s_j} \nonumber \\  
		&+\!\!\left(\!\frac{\eta_{t}L\eta_{t}^2I^2}{2s_{N}} \! +\! \frac{\eta_{t}\gamma_{t}}{2s_{N}}(I \!+\! I^2(1\!+\!\eta_{t})^2)\!+\! L\eta_{t}^2I\!\right)\!\frac{\sum_{n=1}^{N}p_{n,t}s_{n}^2}{\sum_{j=1}^N p_{j,t}s_j}\sigma^2, 
	\end{align}
	\emph{if the following two conditions are satisfied:\vspace{-0.1in}} 
	\begin{align}
		\label{con1}
		&L^2 \eta_{t}^2 I^2 + \gamma_{t}I^2(1+ \eta_{t})^2 + 2 s_{N} L\eta_t I \leq 1,   \\ 
		&\label{con2} L^2 \eta_{t}^2 \gamma_{t} I^2 + \gamma_{t}^2 \eta_{t}^2 I^2 + 2 s_{N} \eta_{t} \gamma_{t}LI \leq 1,
	\end{align}
	\emph{where $p_{n,t} = \exp\left(-\frac{\delta_{n} + BN_{0}}{P_n d_{n}^{-\alpha}}\left(2^{\frac{s\left(\boldsymbol{w}^{(n)}_{t}\right)}{B\left(\bar{t} - I \frac{\bar{s}c}{\phi}\right)}}-1 \right)\right)$}.
	\begin{proof}[Proof:\nopunct]
		The proof is provided in Appendix \ref{prooffortheorem1}.
	\end{proof}
\end{theorem}
Using Theorem \ref{theorem1}, we can calculate how much the total loss decreases between two consecutive communication rounds and determine the speed with which the model converges to the optimal auto-tuning model in (\ref{optimization}).
In particular, as observed from Theorem \ref{theorem1}, the convergence speed hinges on the value of participation probability $p_{n,t}$, $n \in \mathcal{N}$. 
This participation probability depends on the quality of the wireless channel and the distance between the CAVs and the server, as determined by the mobility of the CAVs.
In addition, to identify how the participation of a particular CAV in the FL affects the convergence, we also need to consider the size and distribution of the local data at CAVs. 
To do so, in the following corollary, we will first mathematically define the local data quality of CAVs and then study the impact of local data quality on the convergence of learning models.

\begin{corollary}
	\label{corollary3}
	%\vspace{-0.03in}
	\emph{Given the conditions in (\ref{con1}) and in (\ref{con2}), the local data quality of CAV $n \in \mathcal{N}$ can be defined as 
		\begin{align}
			\beta_{n}\!=&s_{n}^{2}\Bigg[\!\left(\frac{\eta_{t}}{2s_{N}}\!+\!\frac{\gamma_{t}\eta_{t}}{2s_{N}}\right)I ||\nabla f_{n}(\boldsymbol{w}_{t})||_2^{2}  \nonumber \\ &-\!\left(\frac{\eta_{t}}{2s_{N}}L\eta_{t}^2I^2  \!+\! L\eta_{t}^2I\right) \sigma^2 +\! \frac{\eta_{t}\gamma_{t}}{2s_{N}}(I \!+\! I^2(1\!+\!\eta_{t})^2)\sigma^2\Bigg]. \nonumber 
		\end{align}
		The set $\mathcal{N}$ can be divided into two subsets $\mathcal{N}_{(1)}$ and $\mathcal{N}_{(2)}$ with the negative and positive data quality, respectively. 
		In this case, the results in (\ref{major1}) can be simplified as\vspace{0.01in}}
	\begin{align}
		\label{exp_corollary3}
		f(\boldsymbol{w}_{t})\!-\!\mathbb{E}_{\xi, n}(f(\boldsymbol{w}_{t+1}))\geq \frac{\sum_{n \in \mathcal{N}_{(1)}}p_{n,t}\beta_{n}}{\sum_{j=1}^{N}p_{j,t}s_j} \!+\!\sum_{n \in \mathcal{N}_{(2)}}\frac{p_{n,t}\beta_{n}}{s_{N}}. 
	\end{align}
	\begin{proof}[Proof:\nopunct]
		The proof is provided in Appendix \ref{proofforcorollary3}.
		%\vspace{-0.03in}
	\end{proof}
\end{corollary}
According to Corollary \ref{corollary3}, the local data quality for a CAV $n\in\mathcal{N}$ can be calculated based on the size $s_{n}$ of its local data samples and the loss function $ f_{n}(\boldsymbol{w}_{t})$. Also, from Corollary \ref{corollary3}, we observe that the participation of CAVs within the subset $\mathcal{N}_{(1)}$ in the FL will impede the convergence whereas the participation of CAVs from subset $\mathcal{N}_{(2)}$ will improve the FL convergence. 
In other words, depending on the value of the data quality $\beta_{n}, n \in \mathcal{N}$, the convergence gain contributed by different CAVs can be negative or positive.
Note that different from the previous works in \cite{konevcny2016federated}, \cite{9210812}, and \cite{li2020federated}, we mathematically capture the local data quality and analyze the impact of diverse data quality on convergence.
In the following corollary, we also extend Theorem \ref{theorem1} to the case in which the vanilla FedAvg is used for the autonomous controller design.
\begin{corollary}
	\label{corollary1}
	\emph{When using FedAvg algorithm, i.e., no $L_2$ regularizer in each SGD, we can obtain the following upper bound for the expected loss:} 
	\begin{align}
		\mathbb{E}_{\xi, n}(f(\boldsymbol{w}_{t+1}))\!\leq\!&f(\boldsymbol{w}_{t}) \!-\!\frac{\eta_{t}}{2s_{N}}\frac{\sum_{n=1}^{N}p_{n,t}s_n^2I||\nabla f_{n}(\boldsymbol{w}_{t})||_2^{2}}{\sum_{j=1}^{N}p_{j,t}s_j}\!\nonumber \\ &+\!\left(\frac{\eta_{t}}{2s_{N}}L\eta_{t}^2I^2\!+\!L\eta_{t}^2I\right)\frac{\sum_{n=1}^{N}p_{n,t}s_{n}^2}{\sum_{j=1}^N p_{j,t}s_j}\sigma^2,  \nonumber 
	\end{align}
	\emph{if $L^2 \eta_{t}^2 I^2 + 2 s_{N} L\eta_t I \leq 1$.}
	\begin{proof}[Proof:\nopunct]
		We can replace $\gamma_{t}=0$ in Theorem \ref{theorem1} to obtain the bound.
	\end{proof}
\end{corollary}
By comparing Theorem \ref{theorem1} and Corollary \ref{corollary1}, we can prove that, when the constraint (\ref{con1}) is satisfied, the proposed DFP algorithm can achieve a smaller upper bound for the expected loss than FedAvg.
In other words, the proposed DFP can achieve a faster convergence for the controller design in comparison to the FedAvg algorithm, leading to a fast adaptation to the traffic dynamics for CAVs.
To minimize the energy spent on model training, CAVs can also dynamically adjust the number of iteration $I_n, n\in \mathcal{N},$ of the local SGD performed at each communication round. 
In this case, we can simplify the results in Theorem \ref{theorem1} and obtain
\begin{align}
	\label{12}
	&f(\boldsymbol{w}_{t}) - \mathbb{E}_{\xi, n}(f(\boldsymbol{w}_{t+1})) \geq \nonumber \\  &\frac{\sum_{n=1}^{N}p_{n,t}s_n^2\left[-\left(\frac{\eta_{t}\gamma_{t}}{2s_{N}}(1+\eta_{t})^2 \sigma^2 +\frac{\eta_{t}}{2s_{N}}L\eta_{t}^2\sigma^2  \right)I_n^2  \right]}{\sum_{j=1}^N p_{j,t}s_j} + 
	\nonumber \\ 
	&\frac{\sum_{n=1}^{N}p_{n,t}s_n^2\!\!\left[\!\left(\!\left(\!\frac{\eta_{t}}{2s_{N}}\!\!+\!\!\frac{\gamma_{t}\eta_{t}}{2s_{N}}\!\right) \!||\nabla f_{n}(\boldsymbol{w}_{t})||_2^{2} \!\!-\!\!\frac{\eta_{t}\gamma_t\sigma^2}{2s_{N}}\!\! -\!\! L\eta_{t}^2 \sigma^{2}\!\right)I_n\! \right]}{\sum_{j=1}^N p_{j,t}s_j}. 
	%\vspace{-0.04in}
\end{align} 
The result in (\ref{12}) is useful for applications with stringent energy constraints, such as electric CAVs. 
Also, (\ref{12}) can provide guidelines on how to choose the number of local SGD iterations at each CAV so as to facilitate the  convergence to the optimal controller model.

In summary, in this section, we designed the DFP algorithm to tackle the challenges of non-IID and unbalanced data and varying participation of CAVs in the learning process when using FL for the autonomous controller. 
We further proved the convergence and theoretically studied how the data quality, mobility, wireless fading channels, and number of local training iterations affect the overall convergence. 
Based on these insights, next, we will design a contract-theory based incentive mechanism to further improve the convergence performance of the proposed DFP algorithm.

\section{Contract-Theory Based Incentive Mechanism Design}
To improve the controller convergence performance, one can design the incentive mechanism which motivates the CAVs with positive $\beta$ to participate in FL and prevents CAVs with negative $\beta$ from engaging in the FL process.
However, due to the information asymmetry between the server and the CAVs, the server cannot obtain the needed information on the distribution of the local data at each CAV, let alone the data quality. 
To address such information asymmetry, we use the framework of contract theory \cite{bolton2005contract} to design an efficient incentive mechanism for the FL-based autonomous controller design where the parameter server and CAVs are modeled as, respectively, employer and employees in a labor market. 
Contract theory is apropos here because the parameter server can avoid iterative communications with CAVs and increase its utility by allowing the CAVs to instantly choose from a limited number of designed contracts. 
	
There are many conventional approaches to design the incentive mechanism, but unlike the proposed contract-theoretic approach, they are not suitable for the CAV controller design. 
For example, when using the deep reinforcement learning approach \cite{8963610}, it will take a long time to converge to an effective incentive mechanism, inevitably delaying the controller training process and jeopardizing the CAVs’ operation.
Moreover, another alternative approach is to use a Stackelberg game \cite{9247530}. However, in a game setting, each CAV will seek to maximize its own individual utility and, thus, such a strategy may not maximize the parameter server's utility as done in the proposed contract-based approach.
As will be evident from the discussion below, the utility at the parameter server is modeled as the convergence of the learning process, and maximizing the utility at the server is the key goal of our problem.
Hence, to avoid a long delay and improve the FL convergence, we prefer to use contract theory over other alternatives. %}
In the designed contract, the parameter server groups CAVs into different types  according to the data quality $\beta_{n}, n \in \mathcal{N}$, and then designs a unique contract for each type of CAVs. 
In this case, when faced with a list of contracts offered by the parameter server, each CAV will self-reveal the type of its local data quality by choosing the contract designed for its type.
Since the data quality is contingent on how CAVs impact the FL convergence, the designed contract can improve the convergence of the FL-based controller to the optimal CAV controller. 
Next, we will define the utility functions for the parameter server and CAVs and design the contract for the FL-based autonomous controller design. 
\subsection{Utility Function of the Parameter Server}
From Corollary \ref{corollary3}, we can obtain a modified data quality as $\theta_{n}=\frac{\beta_{n}}{s_{N}}, n \in \mathcal{N}$. 
Based on the modified data quality, we assume that all CAVs in set $\mathcal{N}_{(2)}$ can be categorized into $M$ types sorted in an ascending order: $0<\theta_{1}\leq ... \leq \theta_{M}$.
For CAVs in the set $\mathcal{N}_{(1)}$, their corresponding type is denoted as type $0$ with $\theta_{0}=0$.
Clearly, for CAVs belonging to a higher type, their data quality is better and their participation in the FL can expedite the convergence to the optimal autonomous controller model used by CAVs. 
While the parameter server cannot identify the type of a CAV $n\in\mathcal{N}$, we assume that the parameter server has the knowledge of the probability $\bar{p}_{m}$ that a CAV belongs to type $m\in \{1,...,M\}$ based on the historical data and previous observations, as considered in \cite{8964354,9057543,8851649}.

To achieve the self-revealing property, the parameter server will design the contract, i.e., the resource-reward bundle, for each type of CAVs. 
In particular, to compensate the energy consumption spent on the uplink transmission and local training, the resource-reward bundle for CAVs of type $m \in \{1,...,M\}$ can be written as $(P_m, R_m)$, where $R_{m}$ is the reward to the CAVs with an uplink transmit power $P_m$. 
Since CAVs belonging to subset $\mathcal{N}_{(1)}$ actually impede the FL convergence, the parameter server will not give them any compensation, i.e., $R_{0}=0$.
This zero compensation can result in the unwillingness of those CAVs to participate in the FL process, leading to $P_{0}=0$.
However, when incentivizing CAV $n\in \mathcal{N}_{(2)}$ of type $m\in \{1,...,M\}$ into FL aggregation, the utility function of the parameter server at communication round $t$ will be
\begin{align}
	U_{\textrm{ps}}(m)\!\! =\!\! u_1  \exp\left(\!-\frac{\delta_{n}\! \!+\!\! BN_{0}}{P_m d_{n}^{-\alpha}}\left(\!2^{\frac{s\left(\boldsymbol{w}^{(n)}_{t}\right)}{B\left(\bar{t} \! -\! I \frac{\bar{s}c}{\phi}\right)}}\!-\!1 \!\right)\!\right) \theta_m - u_2 R_m, \nonumber 
\end{align}
where $u_1$ captures the valuation factor for the convergence gain brought by the participation of CAVs and $u_2$ is the unit cost of providing a reward to the CAVs. 
As the CAVs in $\mathcal{N}_{(1)}$ are sorted into type $0$ with the reward as $R_{0}=0$ and the transmit power as $P_0=0$, the average utility for the parameter server at communication round $t$ can be written as 
\begin{align}
	\label{utility_ps}
	U_{\textrm{ps}} \!\!= \!\!\!\sum_{n=1}^{N}\!\!\sum_{m=1}^{M}\!\bar{p}_{m}\!\!\Bigg( \!\!u_1 \! \exp\!\Big(\!\!-\!\!\frac{\delta_{n}\!\! + \!\!BN_{0}}{P_m d_{n}^{-\alpha}}\big(2^{\frac{s\left(\boldsymbol{w}^{(n)}_{t}\right)}{B\left(\bar{t} \!-\! I \frac{\bar{s}c}{\phi}\right)}}\!\!-\!\!1 \big)\!\Big) \theta_m\!\! -\!\! u_2 R_m\!\!\Bigg).  
\end{align} 
\vspace{-0.1in}
\subsection{Utility Function of the CAVs}
For the CAVs, the reward received from the parameter server will be used to compensate the energy consumption spent on local model training and uplink transmission.
The utility of CAVs of type $m \in \{1,...,M\}$ is thereby obtained as
\begin{align}
\label{utility_CAV}
U_{\textrm{CAV}}(m) = \theta_{m}R_m - u_{3}(\kappa c \phi^2 \bar{s} I+P_m \hat{t}),
\end{align} 
where $u_{3}$ is the unit cost of the energy consumption.
\subsection{Contract Design}
\label{IIIC}
With the utility functions obtained in (\ref{utility_ps}) and (\ref{utility_CAV}), respectively, for the parameter server and CAVs, we can design the optimal contract which can maximize the utility, i.e., the convergence gain between two consecutive communication rounds, at the parameter server.
In particular, to design a feasible contract for the autonomous controllers, two constraints must be satisfied. 
First, the designed contract must meet the individual rationality (IR) constraint where every CAV is rational and will not accept the contract with a negative utility \cite{bolton2005contract}. That is,
\begin{align}
\label{IRrequirement}
U_{\textrm{CAV}}(m) = \theta_{m}R_m \!-\! u_{3}(\kappa c \phi^2 \bar{s} I\!+\!P_m \hat{t}) \geq 0, m \in \{1,...,M\}. 
\end{align}
For CAVs in type $0$, since $R_{0}=0$, the CAVs will not train their local controller model and will not participate in the uplink transmission, justifying $P_{0}=0$.
Moreover, for a feasible contract, we must impose an incentive compatibility (IC) constraint ensuring that each type of CAVs must always prefer to choose the contract designed for their type over contracts for other types \cite{bolton2005contract}.
In particular, the IC constraints of contract types $m$ and $\hat{m} , \forall m, \hat{m} \in \{1,...,M\}$, will be
\begin{align}
\label{ICrequirement}
\theta_{m}R_m - u_{3}(\kappa c \phi^2 \bar{s} I&+P_m \hat{t}) \geq \theta_{m}R_{\hat{m}} - u_{3}(\kappa c \phi^2 \bar{s} I+P_{\hat{m}} \hat{t}), \nonumber \\ &\forall m, \hat{m} \in \{1,...,M\}. 
\end{align}  

According to (\ref{IRrequirement}) and (\ref{ICrequirement}), we can further simplify the IR and IC constraints and obtain the list of five following conditions for a feasible contract. 

\begin{lemma}
\label{lemma_feasible}
\emph{The designed contract $(P_m, R_m), m \in \{1,...,M\},$ will be feasible if and only if the following five conditions are satisfied:}
\begin{align}
&\text{\textbullet} \hspace{0.05in}\sum_{m=1}^{M}\bar{p}_{m}R_{M} \leq R_{\emph{total}},\label{cond4}\\
& \text{\textbullet}  \hspace{0.05in} 0 \leq R_{1} \leq ... \leq R_{m} \leq ... \leq  R_{M},  \nonumber \\  &\hspace{0.15in}0 \leq P_{1} \leq ... \leq P_{m} \leq ... \leq  P_{M} \leq P_{\max}\label{cond5} \\ 
%&\text{\textbullet} , \label{cond1}\\
&\text{\textbullet} \hspace{0.05in} \theta_{1}R_{1}-u_{3}(\kappa c \phi^2 \bar{s} I + P_{1}\hat{t}) \geq 0, \label{cond2}\\
&\text{\textbullet} \hspace{0.05in} \theta_{m-1}(R_{m}\!-\!R_{m-1})  \!\leq\! u_{3}\hat{t}(P_{m}\!-\!P_{m-1}) \leq \theta_{m}(R_{m}\!-\!R_{m-1}),\nonumber \\  &\hspace{0.15in} m \in \{1,...,M\}\label{cond3},
\end{align}
\emph{where $R_{\textrm{total}}$ is total reward at the parameter server and $P_{\max}$ denotes the maximum transmit power of CAVs.}
\end{lemma}
The condition in (\ref{cond4}) stems from the fact that the parameter server has a limited reward to offer in a contract. 
The proofs for conditions in (\ref{cond5})-(\ref{cond3}) are similar to \cite{bolton2005contract}. 
Based on the utility function defined in (\ref{utility_ps}) and the conditions presented in Lemma \ref{lemma_feasible}, we can formulate the contract design into an optimization problem whose goal is to maximize the average utility at the parameter server, as follows: 
\begin{align}
\label{contract_opt}
&\max_{(P_m,R_m)_{1\!\leq\! m\! \leq\! M}} \sum_{n=1}^{N}\!\sum_{m=1}^{M}\!\bar{p}_{m}\!\left(\!\! u_1 \! \exp\!\left(\!\!\frac{-\!(\!\delta_{n}\! \!+ \!\!BN_{0}\!)A_{n}}{P_m d_{n}^{-\alpha}}\!\right)\! \theta_m \!\!-\!\! u_2 R_m\!\right) \\ 
&\text{s.t.} \hspace{0.1in} (\ref{cond4}), (\ref{cond5}), (\ref{cond2}), (\ref{cond3}), \nonumber 
\end{align}
where $A_{n}=\Big(2^{\frac{s(\boldsymbol{w}^{(n)}_{t})}{B\left(\bar{t} - I \frac{\bar{s}c}{\phi}\right)}}-1 \Big)$.
Due to the non-concave objective function and the complex constraints, directly solving the optimization problem in (\ref{cond4})-(\ref{contract_opt}) will be challenging. 
Alternatively, we will use a sequential method where the optimal power allocation is first  determined in terms of the reward assignment and the optimal reward assignment for each data quality type is then derived.
In the following theorem, we will study the optimal power allocation when the reward assignment is given. 

\begin{theorem}
\label{theorem2}
\emph{Given a reward assignment $\boldsymbol{R}=(R_{1},...,R_{M})$ that satisfies conditions (\ref{cond4}) and (\ref{cond5}), the power allocation $\boldsymbol{P}^{*}=(P^{*}_{1},...,R^{*}_{M})$ that maximizes the average utility at the parameter server will be} 
\begin{align}
\label{sol1}
P^{*}_{m}= \frac{\theta_{1}R_{1}-u_{3}\kappa c \phi^{2}\bar{s}I}{u_{3}\hat{t}} + \sum_{k=1}^{m}  \rho_{k}, m \in \{1,...,M\},
\end{align} 
\emph{where $\rho_{k} = 
0$, if $k =1$; otherwise, $\rho_{k}=\frac{\theta_{k} (R_{k}-R_{k-1})}{u_{3}\hat{t}}$.}
\begin{proof}[Proof:\nopunct]
To prove the optimality of the solutions in (\ref{sol1}), we will proceed by contradiction. 
In particular, we assume there exists another feasible contract $(\boldsymbol{P}', \boldsymbol{R})$ which achieves a higher average utility for the parameter server than the contract $(\boldsymbol{P}^{*}, \boldsymbol{R})$.
Since the utility function at the parameter server is an increasing function of the transmit power, there will be at least one type, e.g., type $\hat{m} \in \{1,...,M\}$, of CAVs with $P^{'}_{\hat{m}}> P^{*}_{\hat{m}}$.
Here, we consider two cases with $\hat{m}=1$ and $\hat{m}\neq 1$. 
When $\hat{m}=1$, $P'_{1} > P^{*}_{1}$. 
As defined in (\ref{sol1}), $\theta_{1}R_{1}-u_{3}(\kappa c \phi^2 \bar{s} I + P^{*}_{1}\hat{t}) = 0$. 
When the CAVs belonging to type 1 are assigned to power $P'_{1} > P^{*}_{1}$, $\theta_{1}R_{1}-u_{3}(\kappa c \phi^2 \bar{s} I + P'_{1}\hat{t}) < 0$, violating the contract feasibility condition (\ref{cond2}).
When $\hat{m}\neq 1$, we have $P'_{\hat{m}} > P^{*}_{\hat{m}}$. 
From condition (\ref{cond3}), the feasible contract $(\boldsymbol{P}',\boldsymbol{R})$ will satisfy the following condition: 
\begin{align}
	\label{(1)}
	u_{3}\hat{t}(P'_{\hat{m}}-P'_{\hat{m}-1}) \leq \theta_{\hat{m}}(R_{\hat{m}}-R_{\hat{m}-1}).
\end{align} 
Using the definition of $P^{*}_{m}, m \in \{1,...,M\},$ in (\ref{sol1}), the values of $R_{\hat{m}}$ and $R_{\hat{m}-1}$ will meet 
\begin{align}
	\label{(2)}
	R_{\hat{m}}-R_{\hat{m}-1}= \frac{u_{3}\hat{t}(P^{*}_{\hat{m}}-P^{*}_{\hat{m}-1})}{\theta_{\hat{m}}}.
\end{align}
Based on the result in (\ref{(2)}), we can simplify the results in (\ref{(1)}) and obtain $P'_{\hat{m}}-P^{*}_{\hat{m}} \leq P'_{\hat{m}-1}-P^{*}_{\hat{m}-1}$. 
As $P'_{\hat{m}}\geq P^{*}_{\hat{m}}$, $P'_{\hat{m}-1}\geq P^{*}_{\hat{m}-1}$. 
Iteratively, the transmit power allocated to the type 1 CAVs in $(\boldsymbol{P}',\boldsymbol{R})$ will be less than the one in $(\boldsymbol{P}^{*},\boldsymbol{R})$, i.e., $P'_{1} \leq P^{*}_{1}$, which is proved to violate the basic feasible contract constraint. 
Hence, there will not exist a feasible contract that achieves a better average utility at the parameter server than the contract $(\boldsymbol{P}^{*},\boldsymbol{R})$.
In other words, for a given reward assignment $\boldsymbol{R}$, the power allocation in the optimal contract is calculated in (\ref{sol1}).
\end{proof} 
\end{theorem}
With the optimal power allocation in Theorem \ref{theorem2}, we can verify that the feasible conditions in (\ref{cond5})-(\ref{cond3}) will be automatically satisfied when $P^{*}_{1} \geq  0$ and $P^{*}_{M}\leq P_{\max}$.
Next, we can replace $P_{m}$ with $P^{*}_{m}$, $m\in \{1,...,M\},$ in (\ref{cond4})-(\ref{contract_opt}) and reformulate the optimization problem as follows:
\begin{align}
\label{contract_opt_new}
&\max_{\boldsymbol{R}} \sum_{n=1}^{N}\!\!\sum_{m=1}^{M}\!\bar{p}_{m}\!\left(\!\! u_1  e^{\frac{-(u_{3}\bar{t}(\delta_{n} + BN_{0})d_{n}^{\alpha})A_{n}}{\theta_{1}R_{1}-u_{3}\kappa c \phi^{2}\bar{s}I + \sum_{k=1}^{m} u_{3}\bar{t} \rho_{k} }} \theta_m \!-\! u_2 R_m\right) \\ 
&\hspace{0.05in}\text{s.t.} \hspace{0.1in} R_{1} \geq \frac{u_{3}\kappa c \phi^{2}\bar{s}I}{\theta_{1}}, R_{M} \leq \frac{P_{\max}u_{3}\hat{t} + u_{3}\kappa c \phi^{2}\bar{s}I}{\theta_{M}},\label{32}\\
&\hspace{0.32in}\sum_{m=1}^{M}\bar{p}_{m}R_{M} \leq R_{\textrm{total}},  \label{33}\\ 
&\hspace{0.32in} R_{m} \leq R_{m+1},  m \in \{1,...,M-1\} \label{34}, 
\end{align}
where the constraints in (\ref{32}) result from $P^{*}_{1} \geq  0$ and $P^{*}_{M}\leq P_{\max}$, and the constraint in (\ref{34}) is derived from the feasibility constraint in (\ref{cond5}). 
Define $\boldsymbol{\mathcal{R}}$ as a set of all possible non-negative reward assignments where the constraints in (\ref{32}) are met. 
The Lagrangian dual function will be
\begin{align}
&L(\boldsymbol{R}, \lambda, \boldsymbol{\mu}) = \max_{\boldsymbol{R} \in \boldsymbol{\mathcal{R}}}\sum_{n=1}^{N}\sum_{m=1}^{M}\bar{p}_{m}\Bigg( u_1 \times \nonumber \\ &\exp\left(-\frac{u_{3}\bar{t}(\delta_{n} + BN_{0})d_{n}^{\alpha}}{\theta_{1}R_{1}-u_{3}\kappa c \phi^{2}\bar{s}I + \sum_{k=1}^{m} u_{3}\bar{t} \rho_{k} }A_{n}\right) \theta_m - u_2 R_m\Bigg) \nonumber \\ &+ \lambda(R_{\textrm{total}}-\sum_{m=1}^{M}\bar{p}_{m}R_{M}) + \sum_{m=1}^{M-1}\mu_{m}(R_{m+1}-R_{m}),
\end{align}
where $\lambda$ and $\boldsymbol{\mu}=\{\mu_{1},...,\mu_{M-1}\}$ are the Lagrangian multipliers associated to the inequality constraints (\ref{33}) and (\ref{34}). 
Hence, the dual optimization problem will be 
\begin{align}
&\min_{\lambda, \boldsymbol{\mu}} L(\boldsymbol{R}, \lambda, \boldsymbol{\mu}) \hspace{0.1in} \text{s.t.} \hspace{0.1in} \lambda \geq 0, \boldsymbol{\mu}\succeq \boldsymbol{0}_{1 \times (M-1)}.
\end{align}
As the dual optimization problem is always convex, it can be solved by updating Lagrangian multipliers using basic gradient based algorithms.
Note that, since the objective function in (\ref{contract_opt_new}) is not concave, the solution obtained in the dual optimization problem will be suboptimal.
However, instead of tackling the original problem in (\ref{contract_opt_new})-(\ref{34}) with a high complexity, the parameter server can spend less computation cost and delay when solving the low-complexity dual optimization problem. 
For example, when choosing the ellipsoid method to solve the dual optimization problem, the complexity will be $\mathcal{O}((M)^2 \ln(1/\varepsilon))$ where $\varepsilon$ is the accuracy requirement \cite{boyd2004convex}.
Once the reward assignment is determined, the transmit power allocation in the contract design can be derived using Theorem \ref{theorem2}.
%By using the designed contract, we can not only offer a reward to compensate the energy consumption at local CAVs but also expedite the convergence to using the optimal controller.  
%\end{align}
%The factors impact the contribution of each individual CAV to the learning process include the distance to the parameter, the transmit power, the CPU computing frequency, and the property of local data (i.e., size and distribution). 
\begin{table}[!t]
\centering \caption{Simulation parameters.\vspace{-0.05in}}	\resizebox{9cm}{!}{
\begin{tabular}{|c|c|c|}
	\hline
	\multicolumn{1}{|c|}{\textbf{Parameter}} &\multicolumn{1}{l}{\textbf{Description}} & \multicolumn{1}{|l|}{\textbf{Value}}  \\ \hline
	\multirow{1}{*}{$\eta$} & \multicolumn{1}{l}{Learning rate} & \multicolumn{1}{|l|}{$0.01$ }   \\ \hline
	\multirow{1}{*}{$\gamma$} & \multicolumn{1}{l}{Coefficient for the $L_2$ regularizer} & \multicolumn{1}{|l|}{$0.1$ }   \\ \hline
	\multirow{1}{*}{$I$} & \multicolumn{1}{l}{Iteration number of local SGD} & \multicolumn{1}{|l|}{$20$ }   \\ \hline
	\multirow{1}{*}{$P_{\max}$} & \multicolumn{1}{l}{maximum transmit power} & \multicolumn{1}{|l|}{$1$~W}   \\ \hline
	
	\multirow{1}{*}{$\Delta t$} & \multicolumn{1}{l}{Sampling period} & \multicolumn{1}{|l|}{$1$~s}   \\ \hline	
	
	\multirow{1}{*}{$\bar{t}$} & \multicolumn{1}{l}{Duration of each communication round} & \multicolumn{1}{|l|}{$0.02$~s}   \\ \hline	
	
	\multirow{1}{*}{$\kappa$} & \multicolumn{1}{l}{
		Energy consumption efficiency} & \multicolumn{1}{|l|}{$10^{-28}$ \cite{8434285}}   \\ \hline	
	
	\multirow{1}{*}{$c$} & \multicolumn{1}{l}{
		Number of computing cycles per bit} & \multicolumn{1}{|l|}{$10^3$ \cite{8434285}}   \\ \hline	
	
	\multirow{1}{*}{$\phi$} & \multicolumn{1}{l}{
		Frequency of the CPU} & \multicolumn{1}{|l|}{$10^9$~cycles/s \cite{8434285}}   \\ \hline	
	
	\multirow{1}{*}{$N_0$} & \multicolumn{1}{l}{
		Noise power spectral density} & \multicolumn{1}{|l|}{$-174$~dBm/Hz}   \\ \hline	
	
	\multirow{1}{*}{$B$} & \multicolumn{1}{l}{
		Bandwidth} & \multicolumn{1}{|l|}{$1$~MHz}   \\ \hline	
	
	\multirow{1}{*}{$\bar{s}$} & \multicolumn{1}{l}{
		Size of randomly selected data} & \multicolumn{1}{|l|}{$1,000$~bits}   \\ \hline	
	
	\multirow{1}{*}{$M$} & \multicolumn{1}{l}{
		Total number of CAV types} & \multicolumn{1}{|l|}{7}   \\ \hline	
	
	\multirow{1}{*}{$\alpha$} & \multicolumn{1}{l}{
		Path-loss exponent} & \multicolumn{1}{|l|}{2.5}   \\ \hline	 
	
	\multirow{1}{*}{$R_{\textrm{total}}$} & \multicolumn{1}{l}{
		Total reward at the parameter server} & \multicolumn{1}{|l|}{5.0}   \\ \hline	 
	
\end{tabular}}
\label{tableParametersValues}\vspace{0.1in}
\end{table}

\section{Simulation results}
To evaluate the performance of the proposed DFP algorithm, we use two real datasets: The BDD data \cite{yu2020bdd100k} and the DACT data \cite{10.1145/3152178.3152184}.
The BDD data is a large-scale driving video dataset with extensive annotations for heterogeneous tasks, and such dataset is collected under diverse geographic, environmental, and weather conditions across the United States. 
The DACT data is a collection of trajectories collected in the city of Columbus, Ohio, where each trajectory records more than $10$~minutes of driving data and can be divided into multiple segments annotated by the operating pattern, like speed-up and slow-down. 
In terms of the traffic model, we consider a $2$~km $\times 2$~km square area with $20$~lanes randomly located around the center of the square area. 
When using BDD data and the DACT data, we assume that CAVs are randomly assigned to these $20$~lanes and all the training data is randomly split among CAVs to capture the unbalanced distribution of local data. 
Similar to \cite{8778746}, the CAVs' velocity is determined by the headway distance to the preceding CAVs.
For the auto-tuning unit used by CAVs, we consider an ANN model with two hidden layers. In particular, each hidden layer has eight fully connected neurons where the initial weights are chosen randomly from $[0,1]$ and the mean squared error is used as the loss function.
The values of the parameters used for simulations are summarized in Table \ref{tableParametersValues}.
\begin{figure}[!t]
	\centering 
	%\vspace{-0.05in}
	\subfloat[Harsh brake in a traffic accident.]{%
		\includegraphics[width=2.0in,height=1.5in]{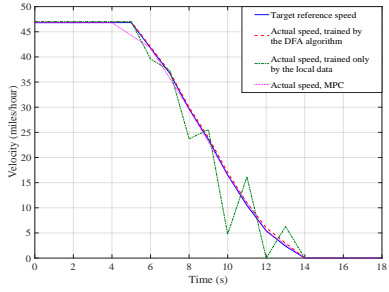}\label{stop}
	}	

	\subfloat[Stop-and-go traffic in a congestion.]{%
		\includegraphics[width=2.0in,height=1.5in]{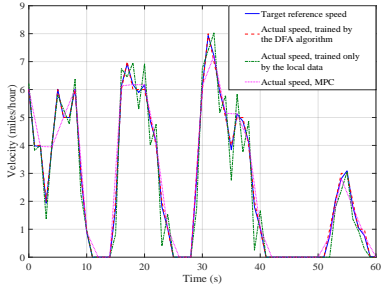}\label{congestion}
	}

	\subfloat[Speed limit changes in a work zone.]{%
		\includegraphics[width=2.0in,height=1.5in]{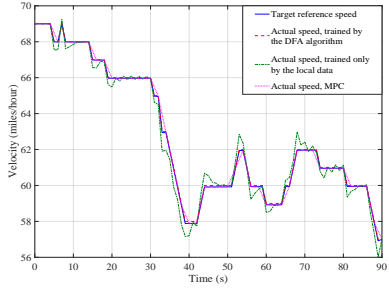}\label{work_zone}
	}
	\caption{Velocity variations over different traffic scenarios.}
	\label{traffic_scenario1}	
	\vspace{-0.05in}
\end{figure}

Fig. \ref{traffic_scenario1} shows the velocity tracking performance comparison between the autonomous controllers
solely trained by the local data (i.e., smooth slow-down) and trained by our proposed DFP algorithm under different traffic scenarios.
In this simulation, we consider three traffic scenarios from the DACT dataset. 
In particular, we choose a use case with a dramatic speed decline to represent a harsh brake in a traffic accident, the speed variations around zero as the stop-and-go traffic in a congestion, and the change of the average speed as the speed limit changes in a roadwork zone. 
As shown in Fig. 2, the controller trained by our proposed DFP algorithm can accurately execute the control decisions and track the target speed under all three traffic scenarios. 
However, when using the controller trained with the local data, we can face large speed variations around the target values. 
For example, as shown in Fig. \ref{traffic_scenario1}\subref{stop}, to achieve a harsh brake, the controller trained by the local data will generate sequential deceleration and acceleration instead of a constant deceleration as done by the controller trained by our proposed DFP algorithm. 
In the traffic congestion and roadwork zone of Figs. \ref{traffic_scenario1}\subref{congestion} and \ref{traffic_scenario1}\subref{work_zone}, the controller trained by the local data will have a more frequent switch between acceleration and deceleration than the target speed traces, adversely impacting the driving experience of the passengers.
Also, in Figs. \ref{traffic_scenario1}\subref{congestion} and \ref{traffic_scenario1}\subref{work_zone}, the controller trained by the local data can make aggressive deceleration and acceleration and such behaviors will not only increase the CAVs' maintenance costs, but it will also endanger following and preceding CAVs especially when the spacing is small.
In Fig. \ref{traffic_scenario1}, we also compare the controller trained by the DFP algorithm with the popular MPC with loss function as the objective function and maximum acceleration constraint as $2.5$~m/s$^2$, and maximum deceleration as $2.5$~m/s$^2$. 
Note that, for MPC, the sampling rate is chosen as $2$~s, due to the fact the MPC needs to solve a quadratic program with a computation complexity higher than the counterparts in our proposed FL based controller design.
In particular, we can observe that, when using the controller trained by our proposed DFP algorithm, the longitudinal velocity trace better aligns with the target reference speed compared to the counterpart of MPC, especially in the stop-and-go traffic. 
Meanwhile, we calculate the mean squared errors for the controller trained by our DFP algorithm which are $0.0993$, $0.0114$, and $0.0032$ for harsh brake, stop-and-go traffic, and speed limit changes. 
For the MPC, the corresponding mean squared errors will be $0.4231$, $0.2751$, and $0.0561$, verifying the effectiveness of our proposed algorithm for the longitudinal controller design.

\begin{figure}
	\centering	
	\begin{minipage}{0.43\textwidth}
		\centering
		\includegraphics[width=0.87\linewidth]{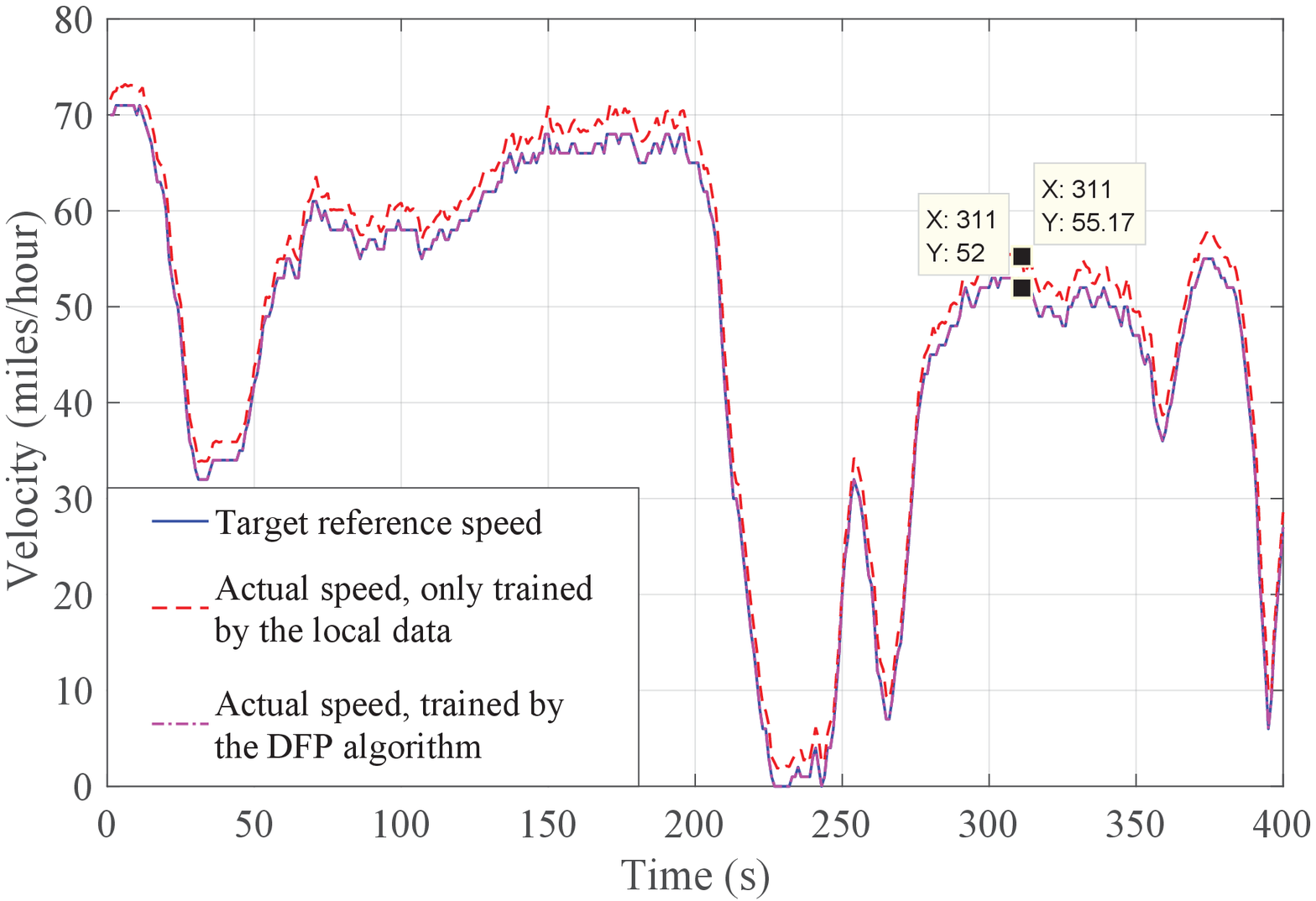}
		\caption{Velocity variations over time.}
		\label{velocity_error}
	\end{minipage}
	\begin{minipage}{0.43\textwidth}
		\centering
		\includegraphics[width=0.87\linewidth]{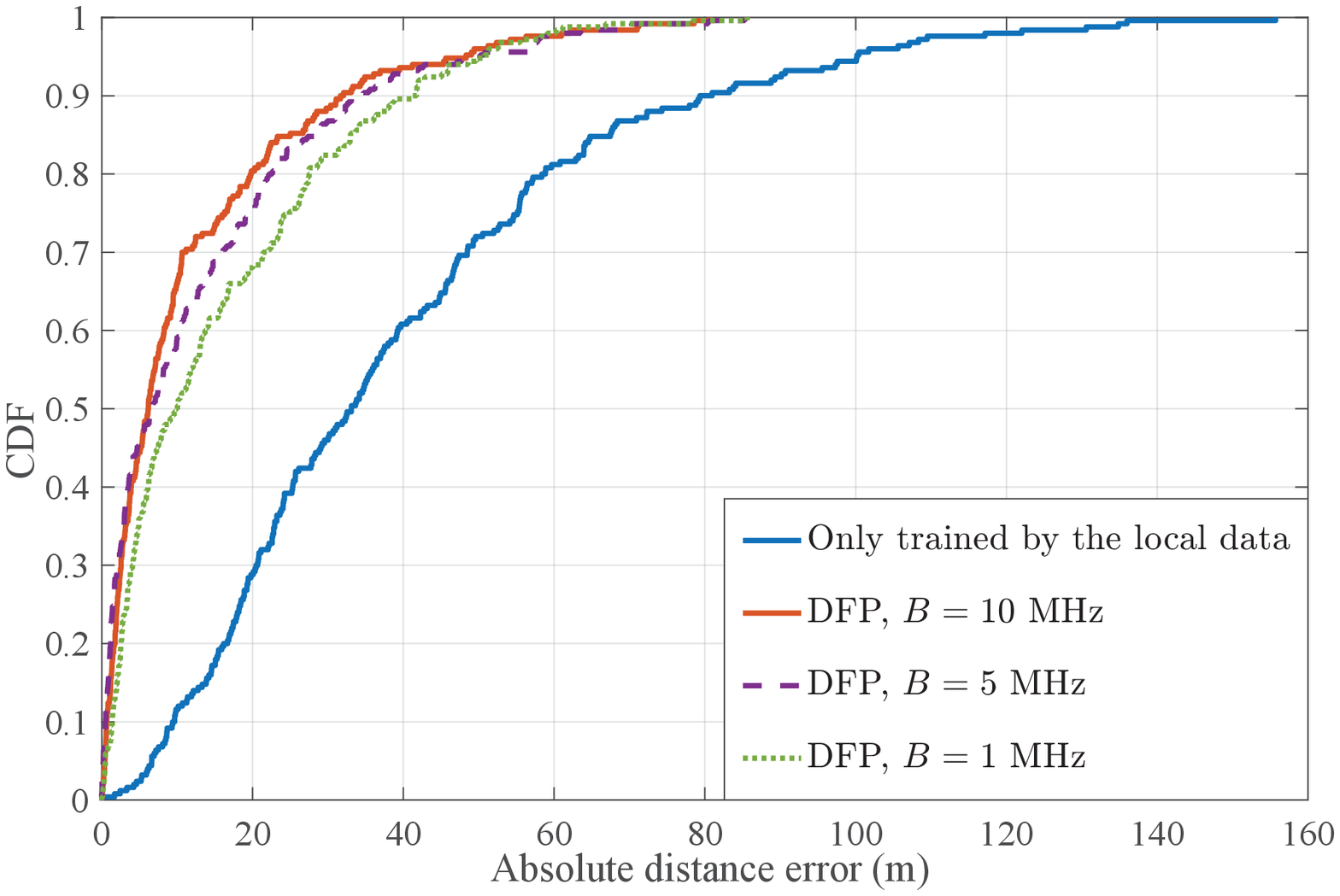}
		\caption{The CDF of absolute distance errors.}
		\label{bandwidth}
	\end{minipage}%%%
	\vspace{-0.05in}
\end{figure}

Fig. \ref{velocity_error} shows the velocity tracking performance comparison between the autonomous controllers solely trained by the local data (i.e., smooth speed-up) and trained by our proposed DFP algorithm over time.
In this simulation, the trajectory data in the DACT dataset is randomly assigned to the CAVs. 
Fig. \ref{velocity_error} shows that the DFP-based controller design can accurately track the target velocity over time. 
However, the actual velocity generated by the controller trained with local data can deviate from the target value.
In particular, at time $t=311$~s, the error between the actual and target velocities can be as large as $3.17$~miles/hour ($1.42$~meters/second), violating the two commonly used design criteria for a vehicle's controller, i.e., $0.5$~meters/second error upper bound \cite{xiong2019speed} and $5\%$ maximum allowable error \cite{6629545}.  

Fig. \ref{bandwidth} shows the cumulative distribution function (CDF) when the controllers tracks the DACT dataset. 
In particular, the autonomous controllers are trained, respectively, by local data and by our proposed DFP algorithm with different bandwidth.
Also, the absolute distance error is calculated by the absolute difference between the target distance in the DACT dataset and the actual distance traversed by the CAV with the designed controller at the end of each trajectory.
As observed from Fig. \ref{bandwidth}, the controller trained by the proposed DFP algorithm yields a much smaller distance error compared with the case in which the CAVs only use their local data to train the controller model. 
In particular, with a $0.90$ probability, the controller solely trained with local data will generate an absolute distance error of less than $80$~m, two times larger than the error resulting from the DFP-based autonomous controller.
Moreover, as shown in Fig. \ref{bandwidth}, for a larger bandwidth, the proposed DFP-based controller design will more likely yield a smaller distance error. 
For example, when the bandwidth $B=10$~MHz, the probability that the distance error generated by DFP-based controller remains below $20$~m is around $0.80$, while the counterpart for the case with a bandwidth $B=1$~MHz is around $0.68$.
That is because with a larger bandwidth, more CAVs can meet the time constraint $\bar{t}$ and participate in the FL, leading to a better training performance. 
As shown in Figures \ref{traffic_scenario1}-\ref{bandwidth}, it is clear that the autonomous controller based on the proposed DFP algorithm outperforms the baseline scheme that solely relies on the local data for training.

\begin{figure}[!t]
	\centering
	\includegraphics[width=2.4in,height=1.6in]{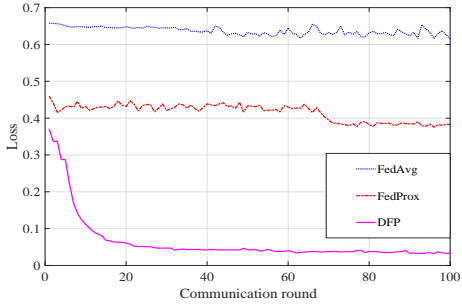}
	\DeclareGraphicsExtensions.
	%\vspace{-0.2cm}
	\caption{\small Convergence performance of the proposed DFP, FedAvg, and FedProx algorithms.}
	\label{comparison}
\end{figure} 

Fig. \ref{comparison} compares the proposed DFP with FedAvg and FedProx. 
To test the ability of dealing with unbalanced and non-IID data for these three algorithms, we choose a larger BDD dataset.
In particular, the BDD data collected under different traffic scenarios will be assigned to different vehicles unevenly to capture the unbalanced and non-IID distribution of local data. 
As observed from Fig. \ref{comparison}, when faced with unbalanced and non-IID training data, FedAvg and FedProx fail to converge near zero loss over $100$~communication rounds. 
In particular, after $100$ communication rounds, the loss values for FedAvg and FedProx are near $0.62$ and $0.38$, respectively.
The slow convergence of FedAvg stems from the fact that the training performance of FedAvg is negatively impacted by the unbalanced and non-IID data. 
The poor performance of FedProx can be explained by the fact that, in FedProx, the CAVs that are randomly selected for the training process might not finish the uplink transmission in time due to the path loss and fading.
However, as shown in Fig. \ref{comparison}, our proposed DFP algorithm needs only around $20$ communication rounds (i.e., $0.2$~s) to achieve convergence, much faster than its counterparts FedAvg and FedProx.
In other words, when dealing with the diverse local data and varying participation of CAVs, our proposed DFP algorithm exhibits a fast convergence to the optimal autonomous controller for CAVs.
Such a fast convergence can enable the CAV to quickly adapt to the traffic dynamics and correctly track the speed determined by the motion planner.

\begin{figure}[!t]
	\centering 
	%\vspace{-0.05in}
	\subfloat[Condition (\ref{cond5}).]{%
		\includegraphics[width=2.9in,height=1.7in]{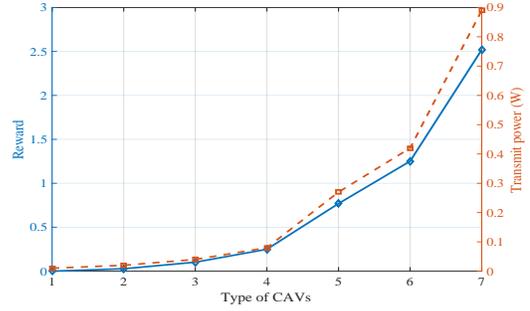}\label{IRcondition}
	}	

	\subfloat[Conditions (\ref{cond2}) and (\ref{cond3}).]{%
		\includegraphics[width=2.9in,height=1.7in]{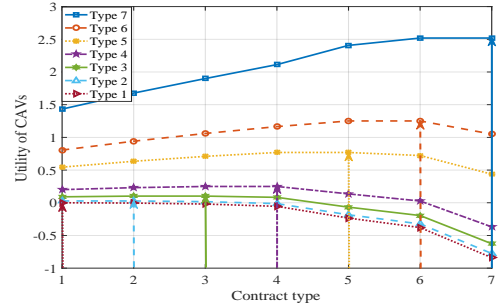}\label{ICcondition}
	}
	\caption{Feasibility of the contract-theory based incentive mechanism for the FL-based controller design.}
	\label{contract_design}	
\end{figure}

In Fig. \ref{contract_design}, we validate the feasibility of the proposed contract-theory based incentive mechanism for the FL-based autonomous controller design among CAVs. 
In particular, as shown in Fig. \ref{contract_design}\subref{IRcondition}, the reward and transmit power increase with the type of CAVs. 
Hence, from Fig. \ref{contract_design}\subref{IRcondition}, our design contract can meet the feasibility constraint (\ref{cond5}). 
Moreover, as shown in Fig. \ref{contract_design}\subref{ICcondition}, we evaluate the utilities of all types of CAVs when selecting all different contracts offered by the parameter server. 
As observed from Fig. \ref{contract_design}\subref{ICcondition}, when choosing contract type 1, the utility of type 1 CAV is non-negative, verifying the feasibility condition (\ref{cond2}). 
Also, we can observe that the utility is a concave function regarding to the CAVs' type, and each type of CAV can achieve its maximum utility if and only if it selects the type of contract that is designed for its own type.
Thus, by using our proposed contract, the CAVs can self-reveal their own types and choose the contract intended for their types in order to maximize the utility, satisfying the feasibility condition (\ref{cond3}).
Hence, given the results in Fig. \ref{contract_design}, we can validate the feasibility of our proposed contract design for the CAVs and parameter server. 

\begin{figure}[!t]
	\centering
	\includegraphics[width=2.7in,height=1.7in]{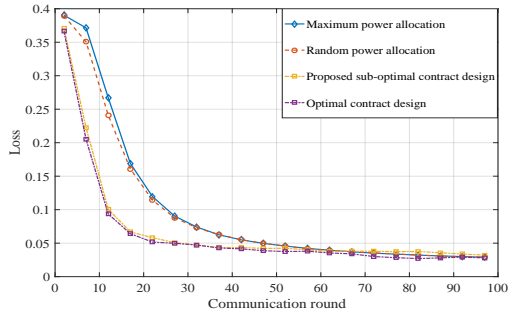}
	\DeclareGraphicsExtensions.
	%\vspace{-0.2cm}
	\vspace{-0.1in}
	\caption{\small Training performance of the proposed contract-based approach and two baselines.}
	% generated by the autonomous controllers trained by local data and our proposed DFP algorithm with different bandwidth.}
\label{merits}
\end{figure}

Fig. \ref{merits} shows the training performance difference when the DFP algorithm uses our proposed  contract-theory based incentive mechanism and two baseline schemes for the power allocation among CAVs. 
The two baseline schemes include the maximum power allocation where all CAVs use the highest transmit power for their uplink transmission and the random power allocation where all CAVs use a randomly selected transmit power in the range from zero to the maximum power. 
In addition, we show the convergence for the optimal contract design where an exhaustive search algorithm is used to determine the optimal reward assignment in (\ref{contract_opt_new}) and then the power allocation in the optimal contract is derived using Theorem \ref{theorem2}.
As shown in Fig. \ref{merits}, we can observe that when using these four assignment strategies, the training loss will decrease as the communication round increases.  
However, the FL process using our proposed contract-theoretic incentive mechanism for the power allocation can achieve a faster convergence compared with random and maximum power allocation schemes. 
In particular, to achieve a $0.05$ loss for the training process of the controller design, the FL process with our proposed scheme will only need around $30$ communication rounds; whereas the corresponding communication rounds for both baseline schemes will be around $50$. 
In other words, our proposed strategy can achieve $40$\% faster FL convergence speed compared with both baseline schemes. 
The reason is that our proposed incentive mechanism will only allocate the transmit power to the CAVs in $\mathcal{N}_{(2)}$ which bring positive convergence gain to the FL process.
However, in the maximum power allocation and the random power allocation, CAVs in $\mathcal{N}_{(1)}$ will also be allowed to participate in the FL and their negative convergence gain will offset the positive gain brought by CAVs in $\mathcal{N}_{(2)}$.
%In this case, when using our proposed incentive mechanism in the autonomous controller design, not only the energy consumption of the CAVs can be compensated but also the CAVs can adapt to the changing traffic dynamics and traffic scenarios more quickly. 
Moreover, we can observe that the convergence of our suboptimal contract design is closely aligned to the optimal contract solution. 
In other words, our suboptimal solution is effective to design a contract which can improve the convergence of the DFP algorithm.
\section{Conclusions}
In this paper, we have developed an FL framework to enable collaborative learning of the autonomous controller model across a group of CAVs. 
In particular, we have proposed a new DFP algorithm that accounts for the varying participation of CAVs in the FL process as well as diverse data quality across CAVs.
We have performed a rigorous theoretical convergence analysis for the proposed algorithm, and we  have explicitly studied the impact of CAVs' mobility, uncertainty of wireless channels, as well as unbalanced and non-IID local data on the overall convergence performance. 
To improve the convergence of the proposed algorithm, we have designed a contract-theoretic incentive mechanism.
Simulation results from using real traces have shown that the autonomous controller designed by the proposed algorithm can track the target speed over time and under different traffic scenarios and the DFP algorithm can lead to a better controller design in comparison to to the FedAvg and FedProx algorithms.  
Also, the simulation results have validated the feasibility of our proposed contract-based incentive mechanism and shown that the incentive mechanism can accelerate the convergence of controller models in CAVs.
In the future work, we will extend the DFP algorithm and the contract theory based incentive mechanism to the stability-oriented adaptive controller design and the scenario with different types of vehicle dynamics.

\appendix
\subsection{Proof of Theorem \ref{theorem1}}

\label{prooffortheorem1}
To prove Theorem \ref{theorem1}, we can find the upper bound of $f(\boldsymbol{w}_{t+1})$ as follows:
{\small\begin{align}
		\label{proof1}
		&f(\boldsymbol{w}_{t+1}) \stackrel{(a)}{\leq} f(\boldsymbol{w}_{t}) + \langle \nabla f(\boldsymbol{w}_{t}), \boldsymbol{w}_{t+1}-\boldsymbol{w}_{t}\rangle \nonumber \\ &+ \frac{1}{2!}(\boldsymbol{w}_{t+1}-\boldsymbol{w}_{t})^T \nabla f(\boldsymbol{w}_{t}) (\boldsymbol{w}_{t+1}-\boldsymbol{w}_{t}) \nonumber \\ 
		&\stackrel{(b)}{\leq} f(\boldsymbol{w}_{t}) + \langle \nabla f(\boldsymbol{w}_{t}), \boldsymbol{w}_{t+1}-\boldsymbol{w}_{t}\rangle + \frac{L}{2}\|\boldsymbol{w}_{t+1}-\boldsymbol{w}_{t}\|_2^2 \nonumber \\
		&\stackrel{(c)}{=}f(\boldsymbol{w}_t)+\Big\langle \sum_{n=1}^{N}\frac{s_n}{s_{N}}\nabla f_{n}(\boldsymbol{w}_t),-\eta_t \sum_{n\in \mathcal{N}_{t}}\frac{s_n}{s_{\mathcal{N}_t}}\sum_{i=0}^{I-1}\Big(\nabla f_{n}(\boldsymbol{w}_{t,i}^{(n)},\xi)\nonumber \\  &+\gamma_{t}(\boldsymbol{w}_{t,i}^{(n)}-\boldsymbol{w}_{t})\Big)\Big\rangle \nonumber \\ 
		&+\frac{L}{2}\left\|\eta_t\sum_{n\in \mathcal{N}_t}\frac{s_n}{s_{\mathcal{N}_{t}}}\sum_{i=0}^{I-1}\left(\nabla f_{n}(\boldsymbol{w}_{t,i}^{(n)},\xi)+\gamma_{t}(\boldsymbol{w}_{t,i}^{(n)}-\boldsymbol{w}_{t})\right)\right\|^2_2 \nonumber \\
		&\stackrel{(d)}{=}f(\boldsymbol{w}_{t}) \!-\! \frac{\eta_{t}}{s_{N}} \sum_{n\in \mathcal{N}_{t}}\frac{s^2_{n}}{s_{\mathcal{N}_{t}}}\sum_{i=0}^{I-1}\left\langle \nabla f_{n}(\boldsymbol{w}_{t}), \nabla f_{n}(\boldsymbol{w}_{t,i}^{(n)},\xi) \right\rangle \nonumber \\&-\!\frac{\eta_{t}\gamma_{t}}{s_{N}} \sum_{n\in \mathcal{N}_{t}}\frac{s^2_{n}}{s_{\mathcal{N}_{t}}}\sum_{i=0}^{I-1}\left\langle \nabla f_{n}(\boldsymbol{w}_{t}), \boldsymbol{w}_{t,i}^{(n)}\!-\!q\boldsymbol{w}_{t} \right\rangle \nonumber \\  &+ \frac{L}{2}\eta_t^2\left\|\sum_{n\in \mathcal{N}_t}\frac{s_n}{s_{\mathcal{N}_{t}}}\sum_{i=0}^{I-1}\left(\nabla f_{n}(\boldsymbol{w}_{t,i}^{(n)},\xi)+\gamma_{t}(\boldsymbol{w}_{t,i}^{(n)}-\boldsymbol{w}_{t})\right)\right\|^2_2,
\end{align}}where \textit{(a)} follows the Taylor expansion, \textit{(b)} is based on the assumption of the Lipschitz continuity, and \textit{(c)} follows the definition of $\nabla f(\boldsymbol{w}_{t})$ and the relationship between $\boldsymbol{w}_{t+1}$ and $\boldsymbol{w}_{t}$. 
And in \textit{(d)}, we use the fact that CAVs train their learning model independently and then further simplify the calculated results. 

In Algorithm \ref{Alg1}, there are two sources of randomness in the FL training. 
First, for $I$ local iterations of SGDs at each communication round, the local data samples selected for training the local FL model will be random.
Second, the CAVs participation in the FL will vary across different communication rounds due to the mobility of CAVs and uncertainty of wireless channels.
We will consider these two sources of randomness sequentially. 
First, when considering the first source of randomness, we take expectation for both sides of  (\ref{proof1}) in terms of the randomly selected set of local samples and we can obtain as follows
{\small\begin{align}
		\label{proof2}
		&\mathbb{E}_{\xi}(f(\boldsymbol{w}_{t+1})) \leq f(\boldsymbol{w}_{t}) \nonumber \\  &- \frac{\eta_{t}}{s_{N}} \sum_{n\in \mathcal{N}_{t}}\frac{s^2_{n}}{s_{\mathcal{N}_{t}}}\sum_{i=0}^{I-1}\mathbb{E}_{\xi}\left\langle \nabla f_{n}(\boldsymbol{w}_{t}), \nabla f_{n}(\boldsymbol{w}_{t,i}^{(n)},\xi) \right\rangle \nonumber \\  &-\frac{\eta_{t}\gamma_{t}}{s_{N}} \sum_{n\in \mathcal{N}_{t}}\frac{s^2_{n}}{s_{\mathcal{N}_{t}}}\sum_{i=0}^{I-1}\mathbb{E}_{\xi}\left\langle \nabla f_{n}(\boldsymbol{w}_{t}), \boldsymbol{w}_{t,i}^{(n)}-\boldsymbol{w}_{t} \right\rangle \nonumber \\ &+ \frac{L}{2}\eta_t^2 \mathbb{E}_{\xi}\left\|\sum_{n\in \mathcal{N}_t}\frac{s_n}{s_{\mathcal{N}_{t}}}\sum_{i=0}^{I-1}\left(\nabla f_{n}(\boldsymbol{w}_{t,i}^{(n)},\xi)+\gamma_{t}(\boldsymbol{w}_{t,i}^{(n)}-\boldsymbol{w}_{t})\right)\right\|^2_2 \nonumber \\ 
		&\stackrel{(a)}{\leq} f(\boldsymbol{w}_{t}) - \frac{\eta_{t}}{s_{N}} \sum_{n\in \mathcal{N}_{t}}\frac{s^2_{n}}{s_{\mathcal{N}_{t}}}\sum_{i=0}^{I-1}\mathbb{E}_{\xi}\left\langle \nabla f_{n}(\boldsymbol{w}_{t}), \nabla f_{n}(\boldsymbol{w}_{t,i}^{(n)}) \right\rangle \nonumber \\ &-\frac{\eta_{t}\gamma_{t}}{s_{N}} \sum_{n\in \mathcal{N}_{t}}\frac{s^2_{n}}{s_{\mathcal{N}_{t}}}\sum_{i=0}^{I-1}\mathbb{E}_{\xi}\left\langle \nabla f_{n}(\boldsymbol{w}_{t}), \boldsymbol{w}_{t,i}^{(n)}-\boldsymbol{w}_{t} \right\rangle \nonumber \\  &+ \frac{L}{2}\eta_t^2 \sum_{n\in \mathcal{N}_t}\frac{s_n^2}{s_{\mathcal{N}_{t}}}\mathbb{E}_{\xi}\left\|\sum_{i=0}^{I-1}\left(\nabla f_{n}(\boldsymbol{w}_{t,i}^{(n)},\xi)+\gamma_{t}(\boldsymbol{w}_{t,i}^{(n)}-\boldsymbol{w}_{t})\right)\right\|^2_2,
\end{align}}where in \textit{(a)}, we use the fact that, for CAV $n\in \mathcal{N}$, the gradient for a random set $\xi \in \mathcal{S}_{n}$ of local data samples is the unbiased estimation to its full gradient representation, i.e., $\mathbb{E}_{\xi} \nabla f_{n}\left(\boldsymbol{w}_{t,i}^{(n)},\xi\right) =\nabla f_{n}\left(\boldsymbol{w}_{t,i}^{(n)}\right)$ \cite{chen2019fast}.
For the second type of randomness, we can take the expectation for both sides of (\ref{proof2}) with respect to the CAVs as follows
{\small\begin{align}
		\label{proof3}
		&\mathbb{E}_{\xi,n}(f(\boldsymbol{w}_{t+1})) \stackrel{(a)}{\leq}\nonumber f(\boldsymbol{w}_{t})  \\ &- \frac{\eta_{t}}{s_{N}} \frac{1}{\sum_{j=1}^{N}p_{j,t}s_{j}} \sum_{n=1}^{N}p_{n,t} s_n^2\sum_{i=0}^{I-1}\mathbb{E}_{\xi,n}\left\langle \nabla f_{n}(\boldsymbol{w}_{t}), \nabla f_{n}(\boldsymbol{w}_{t,i}^{(n)}) \right\rangle \nonumber \\  &-\frac{\eta_{t}\gamma_{t}}{s_{N}}\frac{1}{\sum_{j=1}^{N}p_{j,t}s_{j}} \sum_{n=1}^{N}p_{n,t} s_n^2\sum_{i=0}^{I-1}\mathbb{E}_{\xi,n}\left\langle \nabla f_{n}(\boldsymbol{w}_{t}), \boldsymbol{w}_{t,i}^{(n)}-\boldsymbol{w}_{t} \right\rangle \nonumber \\  &\!+\! \frac{L\eta_t^2}{2} 
		\frac{\sum_{n=1}^{N}p_{n,t} s_n^2
			\mathbb{E}_{\xi,n}\!\left\|\sum_{i=0}^{I\!-\!1}\left(\nabla f_{n}(\boldsymbol{w}_{t,i}^{(n)},\xi)\!\!+\!\!\gamma_{t}(\boldsymbol{w}_{t,i}^{(n)}\!\!-\!\!\boldsymbol{w}_{t})\right)\right\|^2_2}{\sum_{j=1}^{N}p_{j,t}s_{j}}  \nonumber \\ 
		&\stackrel{(b)}{=} f(\boldsymbol{w}_{t}) - \frac{\eta_{t}}{2s_{N}} \frac{1}{\sum_{j=1}^{N}p_{j,t}s_{j}} \sum_{n=1}^{N}p_{n,t} s_n^2\sum_{i=0}^{I-1}\mathbb{E}_{\xi,n}\left\| \nabla f_{n}(\boldsymbol{w}_{t}) \right\|^2_2 \nonumber \\ 
		&- \!\frac{\eta_{t}}{2s_{N}} \frac{\sum_{n=1}^{N}p_{n,t} s_n^2\sum_{i=0}^{I-1}\mathbb{E}_{\xi,n}\left\| \nabla f_{n}(\boldsymbol{w}_{t,i}^{(n)}) \right\|^2_2 }{\sum_{j=1}^{N}p_{j,t}s_{j}}  + \nonumber \\ & \frac{\eta_{t}}{2s_{N}} \frac{\sum_{n=1}^{N}p_{n,t} s_n^2\sum_{i=0}^{I-1}\underbrace{\mathbb{E}_{\xi,n}\left\|\nabla f_{n}(\boldsymbol{w}_{t}) \!- \! \nabla f_{n}(\boldsymbol{w}_{t,i}^{(n)}) \right\|^2_2}_{T_1}}{\sum_{j=1}^{N} p_{j,t}s_{j}}  \nonumber \\ 
		&-\frac{\eta_{t}\gamma_{t}}{2s_{N}}\frac{1}{\sum_{j=1}^{N}p_{j,t}s_{j}} \sum_{n=1}^{N}p_{n,t} s_n^2\sum_{i=0}^{I-1}\mathbb{E}_{\xi,n} \left\|\nabla f_{n}(\boldsymbol{w}_{t}) \right\|^2_2 -\nonumber \\ &\frac{\eta_{t}\gamma_{t}}{2s_{N}}\frac{1}{\sum_{j=1}^{N}p_{j,t}s_{j}} \sum_{n=1}^{N}p_{n,t} s_n^2\sum_{i=0}^{I-1}\mathbb{E}_{\xi,n} \left\|\boldsymbol{w}_{t,i}^{(n)}-\boldsymbol{w}_{t} \right\|^2_2 \nonumber  \\ 
		&+\frac{\eta_{t}\gamma_{t}}{2s_{N}}\frac{\sum_{n=1}^{N}p_{n,t} s_n^2\sum_{i=0}^{I-1}\underbrace{\mathbb{E}_{\xi,n} \left\|\nabla f_{n}(\boldsymbol{w}_{t})-(\boldsymbol{w}_{t,i}^{(n)}-\boldsymbol{w}_{t}) \right\|^2_2}_{T_2}}{\sum_{j=1}^{N}p_{j,t}s_{j}}  \nonumber  \\
		&+\! \frac{L\eta_t^2}{2} 
		\frac{ \sum_{n=1}^{N}p_{n,t} s_n^2
			\underbrace{\mathbb{E}_{\xi,n}\!\left\|\sum_{i=0}^{I-1}\left(\!\nabla f_{n}(\boldsymbol{w}_{t,i}^{(n)},\xi)\!+\!\gamma_{t}(\boldsymbol{w}_{t,i}^{(n)}\!-\!\boldsymbol{w}_{t})\right)\right\|^2_2}_{T_3}}{\sum_{j=1}^{N}p_{j,t}s_{j}},
\end{align}}where in \textit{(a)}, we consider the fact that the probability that CAV $n$ successfully sends its trained model parameters to the parameter server at $t$-th communication round is $p_{n,t}$, and in \textit{(b)}, we follow the fact that for a real vector space, there exists $\langle \boldsymbol{x},\boldsymbol{y} \rangle = \frac{1}{2} (\|\boldsymbol{x}\|_2^2+\|\boldsymbol{y}\|_2^2-\|\boldsymbol{x} -\boldsymbol{y} \|^2_2)$.
In particular, we can simplify $T_{1}$ as follows 
{\small\begin{align}
		\label{proof4}
		&T_1 = \mathbb{E}_{\xi,n}\left\| \nabla f_{n}(\boldsymbol{w}_{t,i}^{(n)})-\nabla f_{n}(\boldsymbol{w}_{t}) \right\|^2_2 \nonumber \\ 
		&\stackrel{(a)}{\leq} L^2 \eta_{t}^2\mathbb{E}_{\xi,n} \left\|\sum_{k=0}^{i-1}\nabla f(\boldsymbol{w}_{t,k}^{(n)},\xi) + \gamma_{t}(\boldsymbol{w}_{t,k}^{(n)}-\boldsymbol{w}_t)\right\|^2_2 \nonumber \\ 
		&\stackrel{(b)}{\leq}2 L^2\eta_{t}^2\mathbb{E}_{\xi,n} \left\|\sum_{k=0}^{i-1}\nabla f(\boldsymbol{w}_{t,k}^{(n)},\xi)\right\|^2_2 \nonumber \\  &+ 2 L^2\eta_{t}^2\mathbb{E}_{\xi,n} \left\|\sum_{k=0}^{i-1} \gamma_{t}(\boldsymbol{w}_{t,k}^{(n)}-\boldsymbol{w}_t)\right\|^2_2 \nonumber \\ 
		&=2 L^2\eta_{t}^2\mathbb{E}_{\xi,n} \left\|\sum_{k=0}^{i-1}\nabla f(\boldsymbol{w}_{t,k}^{(n)},\xi)-\nabla f(\boldsymbol{w}_{t,k}^{(n)})+ \nabla f(\boldsymbol{w}_{t,k}^{(n)}) \right\|^2_2 \nonumber \\  &+ 2 L^2\eta_{t}^2\mathbb{E}_{\xi,n} \left\|\sum_{k=0}^{i-1} \gamma_{t}(\boldsymbol{w}_{t,k}^{(n)}-\boldsymbol{w}_t)\right\|^2_2  \nonumber \\ 
		&=2 L^2\eta_{t}^2\mathbb{E}_{\xi,n} \left\|\sum_{k=0}^{i-1}\nabla f(\boldsymbol{w}_{t,k}^{(n)},\xi)-\nabla f(\boldsymbol{w}_{t,k}^{(n)})\right\|^2_2 \nonumber \\  &+ 2 L^2\eta_{t}^2\mathbb{E}_{\xi,n} \left\|\sum_{k=0}^{i-1}\nabla f(\boldsymbol{w}_{t,k}^{(n)})\right\|^2_2  \nonumber \\ 
		&+ 4L^2\eta_{t}^2\mathbb{E}_{\xi,n}\left \langle\sum_{k=0}^{i-1}\nabla f(\boldsymbol{w}_{t,k}^{(n)},\xi)-\nabla f(\boldsymbol{w}_{t,k}^{(n)}), \nabla f(\boldsymbol{w}_{t,k}^{(n)})\right\rangle \nonumber \\  &+2 L^2\eta_{t}^2\mathbb{E}_{\xi,n} \left\|\sum_{k=0}^{i-1} \gamma_{t}(\boldsymbol{w}_{t,k}^{n}-\boldsymbol{w}_t)\right\|^2_2 \nonumber \\
		&\stackrel{(c)}{\leq}2 L^2\eta_{t}^2 i \sigma^2 + 2 L^2\eta_{t}^2 i\sum_{k=0}^{i-1}\mathbb{E}_{\xi,n} \left\|\nabla f(\boldsymbol{w}_{t,k}^{n})\right\|^2_2 \nonumber \\  &+  2 L^2\eta_{t}^2 \gamma_{t}^2 i \sum_{k=0}^{i-1}\mathbb{E}_{\xi,n} \left\|(\boldsymbol{w}_{t,k}^{n}-\boldsymbol{w}_t)\right\|^2_2,
\end{align}}where \textit{(a)} follows the relationship between $\boldsymbol{w}_{t,i}^{(n)}$ and $\boldsymbol{w}_{t}$, and \textit{(b)} follows the Cauchy-Schwarz inequality $\|1\cdot\boldsymbol{x} +1\cdot \boldsymbol{y} \|_2^2 \leq 2 \|\boldsymbol{x}\|_2^2 + 2 \|\boldsymbol{y}\|_2^2$. 
In \textit{(c)}, we use the fact, for $n\in \mathcal{N}$, $\mathbb{E}_{\xi,n}(\nabla f(\boldsymbol{w}_{t,k}^{(n)},\xi)-\nabla f(\boldsymbol{w}_{t,k}^{(n)}))=0$, the assumption of the Lipschitz continuity, and the extension of Cauchy-Schwarz inequality for $i$ variables, i.e., $\|\sum_{k=0}^{i-1}1\cdot\boldsymbol{x}_{k}\|_{2}^{2} \leq i\sum_{k=0}^{i-1}\|\boldsymbol{x}_{k}\|_{2}^2$. For the term $T_{2}$, we have
{\small\begin{align}
		\label{proof5}
		&T_2 = \mathbb{E}_{\xi,n} \left\|\nabla f_{n}(\boldsymbol{w}_{t})\!-\!\nabla f_{n}(\boldsymbol{w}_{t},\xi )\!+\!\nabla f_{n}(\boldsymbol{w}_{t},\xi ) \!-\!(\boldsymbol{w}_{t,i}^{(n)}-\boldsymbol{w}_{t}) \right\|^2_2 \nonumber \\ 
		&= \!\mathbb{E}_{\xi,n}\! \left\|\nabla f_{n}(\boldsymbol{w}_{t})\!\!-\!\!\nabla f_{n}(\boldsymbol{w}_{t},\xi )\right\|^2_2 \!\!+\! \mathbb{E}_{\xi,n} \!\left\|\nabla f_{n}(\boldsymbol{w}_{t},\xi )\!\! -\!\!(\boldsymbol{w}_{t,i}^{(n)}\!\!-\!\!\boldsymbol{w}_{t}) \right\|^2_2 \nonumber \\
		&+ 2\mathbb{E}_{\xi,n}\left\langle \nabla f_{n}(\boldsymbol{w}_{t})-\nabla f_{n}(\boldsymbol{w}_{t},\xi ), \nabla f_{n}(\boldsymbol{w}_{t},\xi ) -(\boldsymbol{w}_{t,i}^{(n)}-\boldsymbol{w}_{t})\right\rangle \nonumber \\ 
		&\stackrel{(a)}{=}\!\!\mathbb{E}_{\xi,n}\! \left\|\nabla f_{n}(\boldsymbol{w}_{t})\!\!-\!\!\nabla f_{n}(\boldsymbol{w}_{t},\xi )\right\|^2_2 \!\!+\!\! \mathbb{E}_{\xi,n}\! \left\|\nabla f_{n}(\boldsymbol{w}_{t},\xi )\!\! -\!\!(\boldsymbol{w}_{t,i}^{(n)}\!\!-\!\!\boldsymbol{w}_{t}) \right\|^2_2 \nonumber \\
		&\stackrel{(b)}{\leq}\!\!\sigma^2\!\! +\!\! \mathbb{E}_{\xi,n}\left\|\nabla f_{n}(\boldsymbol{w}_{t},\xi )\!\! +\!\! \eta_{t} \left( \sum_{k=0}^{i-1}\nabla f(\boldsymbol{w}_{t,k}^{(n)},\xi)\!\! +\!\! \gamma_{t}(\boldsymbol{w}_{t,k}^{(n)}\!-\!\boldsymbol{w}_t) \right)\right\|^2_2\nonumber \\
		&\stackrel{(c)}{\leq}\sigma^2 + 2\mathbb{E}_{\xi,n}\left\| \nabla f_{n}(\boldsymbol{w}_{t},\xi ) + \eta_{t} \sum_{k=0}^{i-1}\nabla f(\boldsymbol{w}_{t,k}^{(n)},\xi) \right\|^2_2\nonumber \\  &  +2\mathbb{E}_{\xi,n}\left\|\eta_{t} \sum_{k=0}^{i-1}  \gamma_{t}(\boldsymbol{w}_{t,k}^{(n)}-\boldsymbol{w}_t) \right\|^2_2 \nonumber \\ 
		&\stackrel{(d)}{\leq}\!\sigma^2 \!\!+\!\! 2(1\!\!+\eta_{t})^2 \mathbb{E}_{\xi, n}\left\| \sum_{k=0}^{i-1}\nabla f(\boldsymbol{w}_{t,k}^{(n)},\xi)  \right\|^2_2 \nonumber \\ &+ 2\eta_{t}^2\gamma_{t}^2 \mathbb{E}_{\xi,n}\left\| \sum_{k=0}^{i-1} \boldsymbol{w}_{t,k}^{(n)}-\boldsymbol{w}_t\right\|^2_2 \nonumber \\ 
		&\stackrel{(e)}{\leq} \sigma^{2} + 2i(1+\eta_{t})^{2}\sigma^2 + 2(1+\eta_{t})^2 i \sum_{k=0}^{i-1} \mathbb{E}_{\xi, n} \left\|\nabla f(\boldsymbol{w}_{t,k}^{(n)})\right\|^2_2 \nonumber \\  &+  2\eta_{t}^2 \gamma_{t}^2 i\sum_{k=0}^{i-1}\mathbb{E}_{\xi,n} \left\|(\boldsymbol{w}_{t,k}^{(n)}-\boldsymbol{w}_t)\right\|^2_2,
\end{align}}where in \textit{(a)},  we use the fact, for $n\in \mathcal{N}$, $\mathbb{E}_{\xi,n}\left(\nabla f(\boldsymbol{w}_{t,k}^{(n)},\xi)\!-\!\nabla f(\boldsymbol{w}_{t,k}^{(n)})\right)\!=\!0$, and we expand the expression of $\boldsymbol{w}_{t,i}^{(n)}$ in \textit{(b)}. 
Also the changes in \textit{(c)} are based on the fact that $\boldsymbol{w}_{t}\!=\!\boldsymbol{w}_{t,0}^{(n)}$, and in \textit{(d)}, we follow the Cauchy-Schwarz inequality. 
The final inequality in \textit{(e)} exploits the facts used in (\ref{proof4}). 
When considering all $I$ iterations of SGD, $T_{1}$ and $T_{2}$ can be simplified as follows
{\small\begin{align}
		\label{proof7}
		\sum_{i=1}^{I}T_1 \leq& L^2\eta_{t}^2 I^2 \sigma^2 +  L^2\eta_{t}^2 I^2\sum_{i=0}^{I-1}\mathbb{E}_{\xi,n} \left\|\nabla f(\boldsymbol{w}_{t,i}^{(n)})\right\|^2_2 \nonumber \\  &+  L^2\eta_{t}^2I^2 \gamma_{t}^2\sum_{i=0}^{I-1}\mathbb{E}_{\xi,n} \left\|(\boldsymbol{w}_{t,i}^{(n)}-\boldsymbol{w}_t)\right\|^2_2,  \\
		\sum_{i=1}^{I}T_2 \leq& I\sigma^{2} \!\!+\!\! I^2(1+\eta_{t})^{2}\sigma^2 \!\!+\!\! (1+\eta_{t})^2 I^2 \sum_{i=0}^{I-1} \mathbb{E}_{\xi, n} \left\|\nabla f(\boldsymbol{w}_{t,i}^{(n)})\right\|^2_2\nonumber \\  & +  \eta_{t}^2 \gamma_{t}^2 I^2\sum_{i=0}^{I-1}\mathbb{E}_{\xi,n} \left\|(\boldsymbol{w}_{t,i}^{(n)}- \boldsymbol{w}_t)\right\|^2_2.
		\label{proof8}
\end{align}}
To simplify $T_{3}$, we have 
{\small\begin{align}
		\label{proof6}
		&T_{3} \stackrel{(a)}{\leq} 2 \mathbb{E}_{\xi,n}\left\|\sum_{i=0}^{I-1}\nabla f_{n}(\boldsymbol{w}_{t,i}^{(n)},\xi)\right\|^2_2 + 2 \mathbb{E}_{\xi,n}\left\|\sum_{i=0}^{I-1}\gamma_{t}(\boldsymbol{w}_{t,i}^{(n)}-\boldsymbol{w}_{t})\right\|^2_2 \nonumber \\ 
		&\stackrel{(b)}{\leq}2I \sigma^2\! \!+ \!\!2 I \sum_{i=0}^{I-1}\!\mathbb{E}_{\xi,n} \|\nabla f_n(\boldsymbol{w}_{t,i}^{(n)})\|_{2}^{2}\! +\! 2 \gamma_{t}^{2} I\! \sum_{i=0}^{I-1} \mathbb{E}_{\xi,n} \left\|\boldsymbol{w}_{t,i}^{(n)}\!\!-\!\!\boldsymbol{w}_{t}\right\|^2_2,
\end{align}}where \textit{(a)} and \textit{(b)} follow the same facts and laws used in simplifying $T_{1}$ and $T_{2}$.
After replacing (\ref{proof7}), (\ref{proof8}), and (\ref{proof6}) into the corresponding terms in (\ref{proof3}), we can have 
{\small\begin{align}
		\label{proof9}
		&\mathbb{E}_{\xi,n}(f(\boldsymbol{w}_{t+1}))  \leq f(\boldsymbol{w}_{t}) \!\!-\!\!\frac{\left(\frac{\eta_{t}}{2s_{N}}\!\!+\!\!\frac{\gamma_{t}\eta_{t}}{2s_{N}}\right)\sum_{n=1}^{N}p_{n,t}s_j^2I||\nabla f_{n}(\boldsymbol{w}_{t})||_2^{2}}{\sum_{j=1}^{N}p_{j,t}s_j} \nonumber \\ 
		&+\left(\frac{\eta_{t}}{2s_{N}}L\eta_{t}^2I^2 \! +\! \frac{\eta_{t}\gamma_{t}}{2s_{N}}(I + I^2(1+\eta_{t})^2)\!+\! L\eta_{t}^2I\right)\frac{\sum_{n=1}^{N}p_{n,t}s_{n}^2}{\sum_{j=1}^N p_{j,t}s_j}\sigma^2 \nonumber \\ 
		&-\left(\frac{\eta_{t}}{2s_{N}}-\frac{L^2I^2\eta^3_{t}}{2s_{N}} -\frac{\eta_{t}\gamma_{t}I^2(1+\eta_{t})^2}{2s_{N}}-I L\eta_{t}^{2}  \right)\times \nonumber \\ &\frac{\sum_{n=1}^{N}p_{n,t}s_n^2\sum_{i=1}^{I-1}\mathbb{E}_{\xi,n}||\nabla f_{n}(\boldsymbol{w}_{t,i}^{(n)})||_2^{2}}{\sum_{j=1}^{N}p_{j,t}s_j} \nonumber \\ 
		&-\left(\frac{\gamma_{t}\eta_{t}}{2s_{N}}-\frac{L^2I^2\eta^3_{t}\gamma_{t}^{2}}{2s_{N}} -\frac{\eta^{3}_{t}\gamma^{3}_{t}I^2}{2s_{N}}-I L\eta_{t}^{2} \gamma_{t} \right)\times \nonumber \\ &\frac{\sum_{n=1}^{N}p_{n,t}s_n^2\sum_{i=1}^{I-1}\mathbb{E}_{\xi,n}||\boldsymbol{w}_{t,i}^{(n)}-\boldsymbol{w}_{t}||_2^{2}}{\sum_{j=1}^{N}p_{j,t}s_j}. 
\end{align}}By using condition (\ref{con1}) for (\ref{proof9}), we can obtain the results in (\ref{major1}). 
To further simplify the expression of $p_{n,t}$, we have 
{\small\begin{align}
		p_{n,t} &= \mathbb{P}\left(t_{n,\text{comp}}+t_{n,\text{comm}} \leq \bar{t}  \right) \nonumber \\ &=\mathbb{P}\left(I\frac{\bar{s}c}{\phi}+ \frac{s(\boldsymbol{w}^{(n)}_{t})}{B \log_{2} \left(1+\frac{P_n h_{n} d_{n}^{-\alpha}}{\delta_{n} + B N_0}\right)} \leq \bar{t} \right) \nonumber \\ 
		&=\mathbb{P}\left(h_n \geq \frac{\delta_{n} + BN_{0}}{P_n d_{n}^{-\alpha}}\left(2^{\frac{s(\boldsymbol{w}^{(n)}_{t})}{B\left(\bar{t} - I \frac{\bar{s}c}{\phi}\right)}}-1 \right)\right)\nonumber \\  & \stackrel{(a)}{=}\exp\left(-\frac{\delta_{n} + BN_{0}}{P_n d_{n}^{-\alpha}}\left(2^{\frac{s(\boldsymbol{w}^{(n)}_{t})}{B\left(\bar{t} - I \frac{\bar{s}c}{\phi}\right)}}-1 \right)\right), 
\end{align}}where in \textit{(a)}, we use the fact that the channels between the parameter server and the CAVs are Rayleigh fading channels.

\subsection{Proof of Corollary \ref{corollary3}}
\label{proofforcorollary3}
Based on the definition of $\mathcal{N}_{(1)}$ and $\mathcal{N}_{(2)}$, we have 
{\small\begin{align}
f(\boldsymbol{w}_{t})-\mathbb{E}_{\xi, n}(f(\boldsymbol{w}_{t+1}))&\geq \frac{\sum_{n \in \mathcal{N}_{(1)}}p_{n,t}\beta_{n}}{\sum_{j=1}^{N}p_{j,t}s_j} + \frac{\sum_{n \in \mathcal{N}_{(2)}}p_{n,t}\beta_{n}}{\sum_{j=1}^{N}p_{j,t}s_j} 
\nonumber \\  &\stackrel{(a)}{\geq} \frac{\sum_{n \in \mathcal{N}_{(1)}}p_{n,t}\beta_{n}}{\sum_{j=1}^{N}p_{j,t}s_j} + \frac{\sum_{n \in \mathcal{N}_{(2)}}p_{n,t}\beta_{n}}{\sum_{j=1}^{N}s_j},  
\end{align}}where the changes in \textit{(a)} are based on the fact that $\beta_{n}>0$, for $n\in\mathcal{N}_{(2)}$ and the probability term $0 \leq p_{j,t} \leq 1$. 
Since $\sum_{j=1}^{N}s_{j}=s_{N}$, we can obtain the results in Corollary \ref{corollary3}.

\def\baselinestretch{1.00}

\bibliography{references}
\begin{IEEEbiography}[{\includegraphics[width=1in,height=1.25in,clip,keepaspectratio]{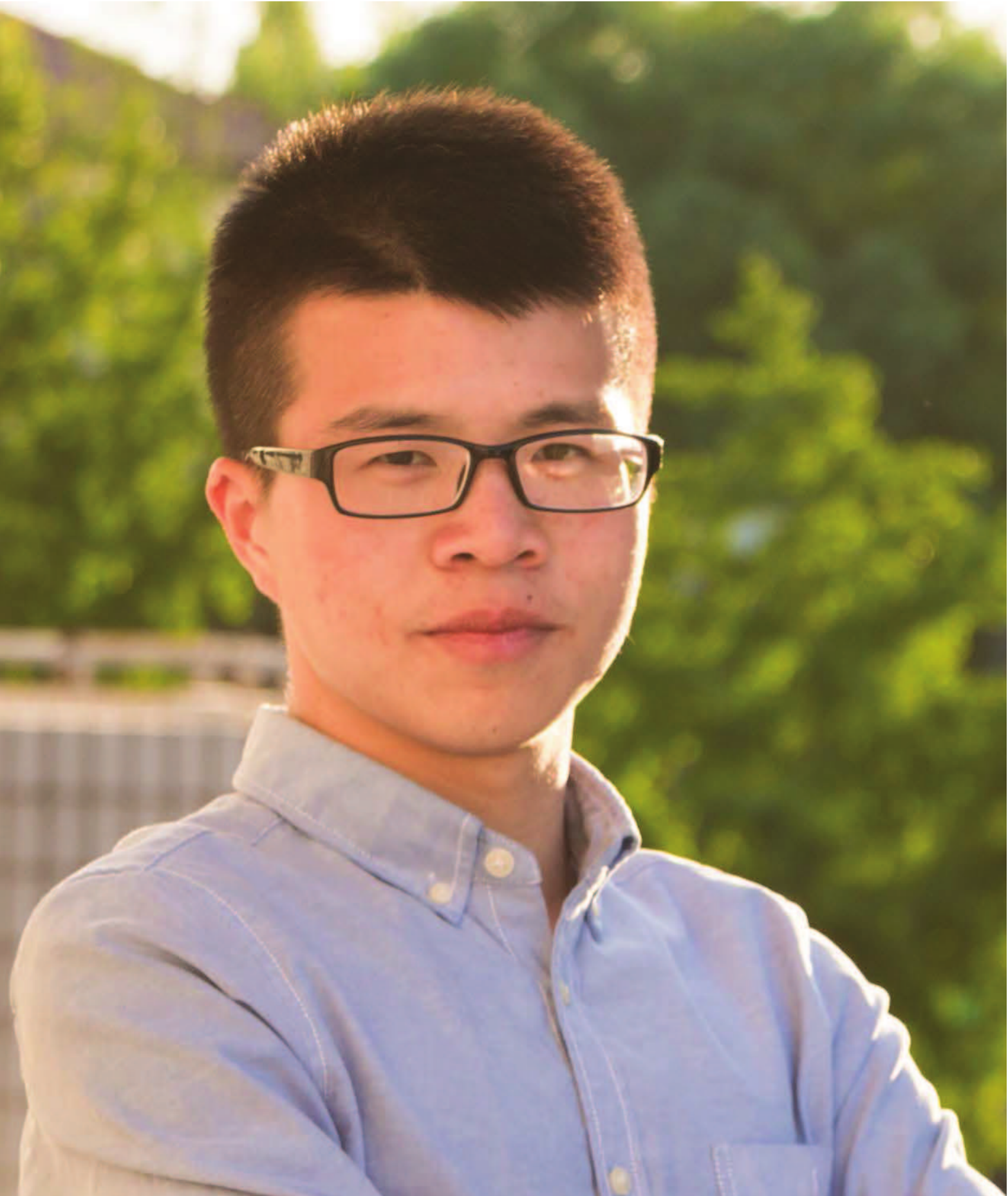}}]{Tengchan Zeng} (S'18) received his BS in Automation (Railway Signal) from the Beijing Jiaotong University in 2016 and his Ph.D. degree from Virginia Tech in 2021. He is currently a Systems Engineer at Ford Motor Company.  
His research interests include control, risk analysis, wireless networks, resource allocation and optimization. He also served as a reviewer for IEEE Transactions and flagship conferences. \vspace{-1cm}
\end{IEEEbiography}

\begin{IEEEbiography}[{\includegraphics[width=1in,height=1.25in,clip,keepaspectratio]{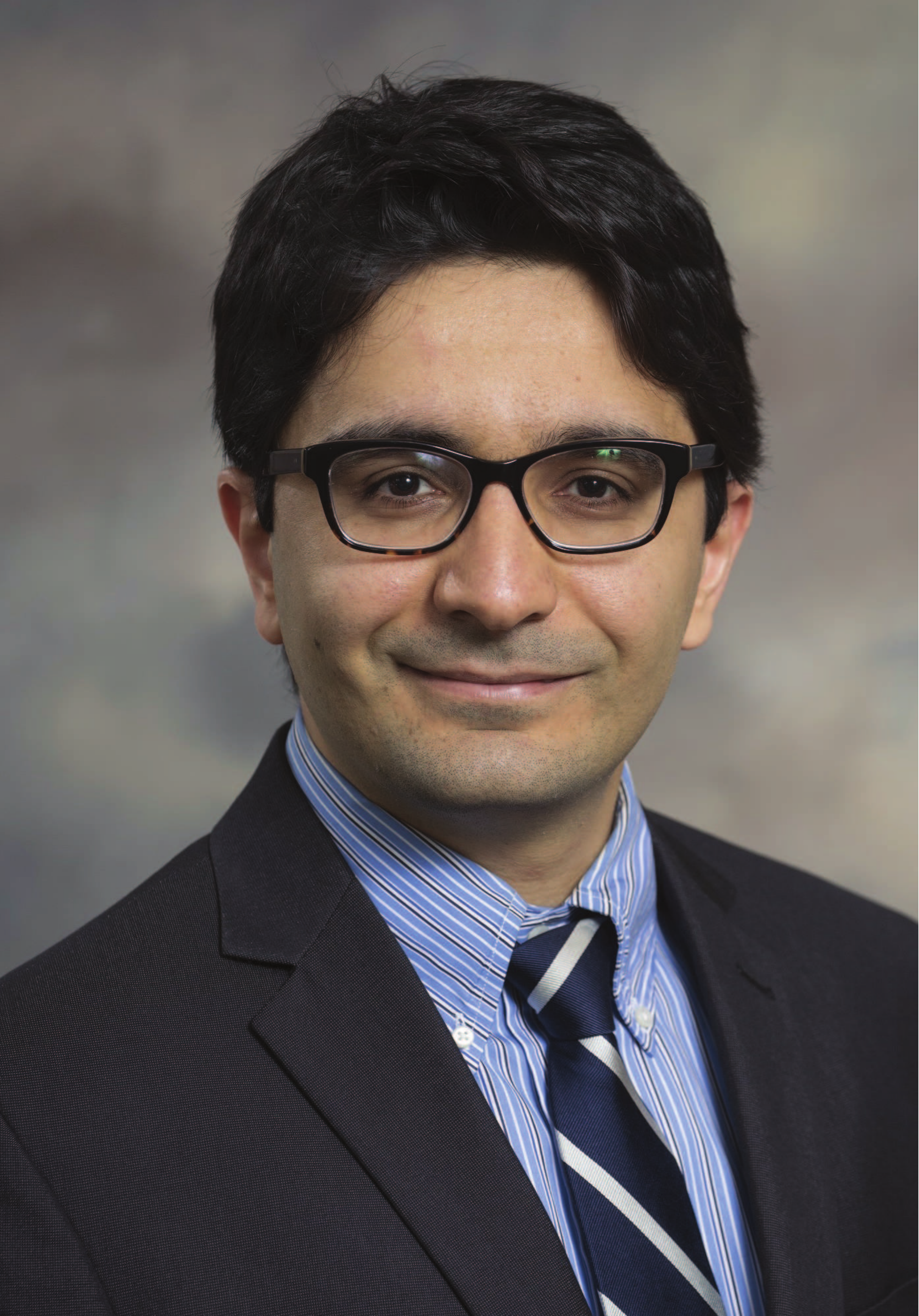}}]{Omid Semiari} (S'14, M'18) received his Ph.D. from Virginia Tech in 2017. He is currently an assistant professor with the Department of Electrical and Com­puter Engineering at the University of Colorado, Colorado Springs. His research interests include wireless communications (6G), machine learning for communications, distributed learning over wireless networks, and cross-layer network optimization, with an emphasis on new technologies such as connected and autonomous vehicles, wireless extended reality, and industrial IoT.  He is the recipient of several research awards, including the NSF CRII award.  He is currently an Associate Editor for the IEEE Journal on Selected Areas in Communications (JSAC), machine learning series. He also serves as an Associate Editor of the Frontiers Journal, a specialty section of Frontiers in Communications and Networks. He has actively served as a reviewer for IEEE Transactions and has been a TPC member for many IEEE flagship conferences.\vspace{-1cm}
\end{IEEEbiography}

\begin{IEEEbiography}[{\includegraphics[width=1in,height=1.25in,clip,keepaspectratio]{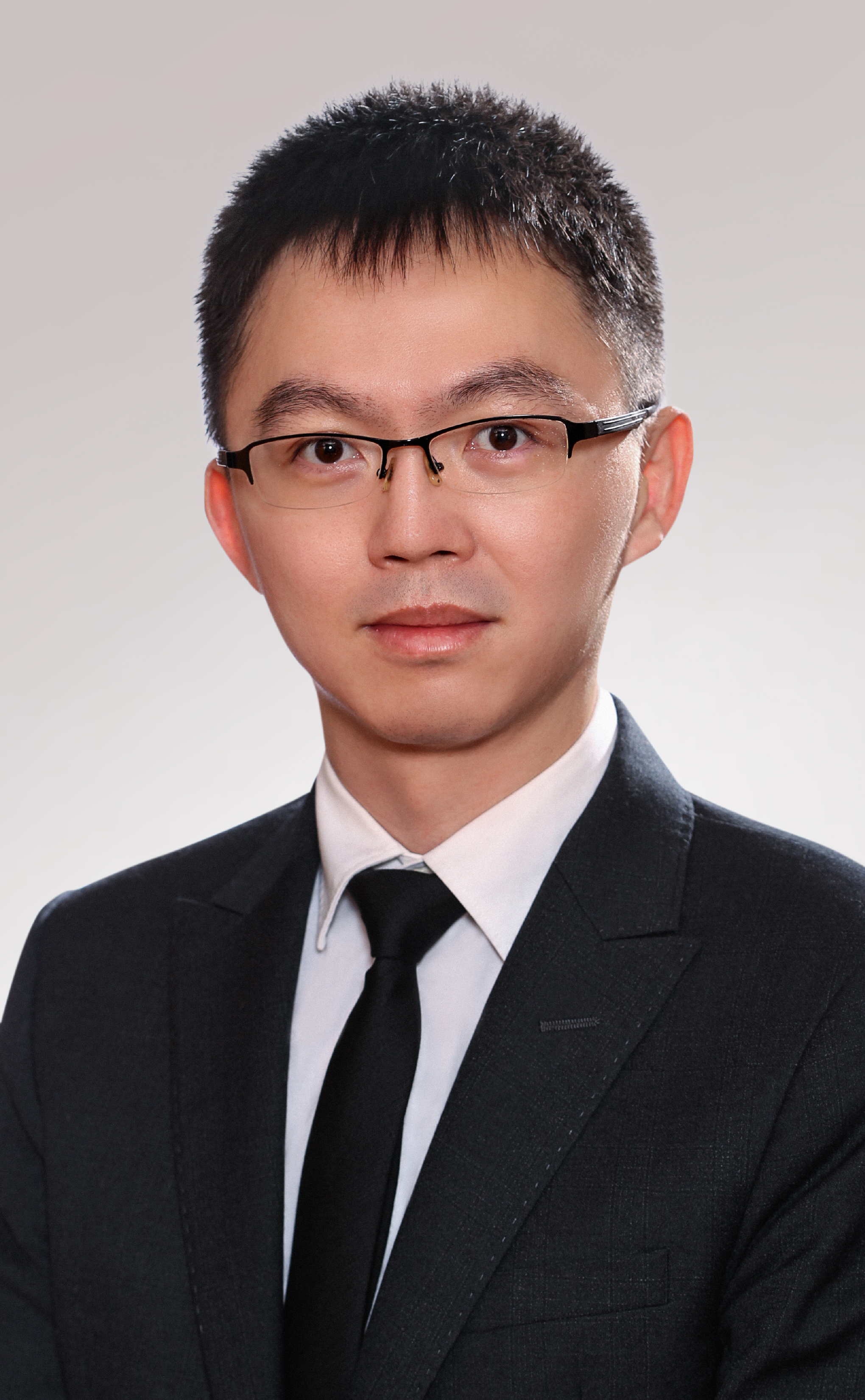}}]{Mingzhe Chen} (S'15, M'19) received his Ph.D. degree from Beijing University of Posts and Telecommunications, Beijing, China, in 2019. From 2016 to 2019, he was a Visiting Researcher at the Department of Electrical and Computer Engineering, Virginia Tech. He is	currently a Postdoctoral Research Associate at the Electrical and Computer Engineering Department, Princeton University. His research interests include federated learning, reinforcement learning, virtual reality, unmanned aerial vehicles, and wireless networks. He is the recipient of the 2021 IEEE ComSoc Young Author Best Paper Award and the 2022	Fred W. Ellersick Prize Award from the IEEE Communications Society. He received three conference best paper awards at IEEE ICC in 2020, IEEE GLOBECOM in 2020, and IEEE WCNC in 2021. He currently serves as an Associate Editor of IEEE Transactions on Green Communications and Networking. Previously, he guest edited a special issue on Distributed Learning over Wireless Edge Networks for IEEE Journal on Selected Areas in Communications (JSAC).\vspace{-1cm}
\end{IEEEbiography}

\begin{IEEEbiography}[{\includegraphics[width=1in,height=1.6in,clip,keepaspectratio]{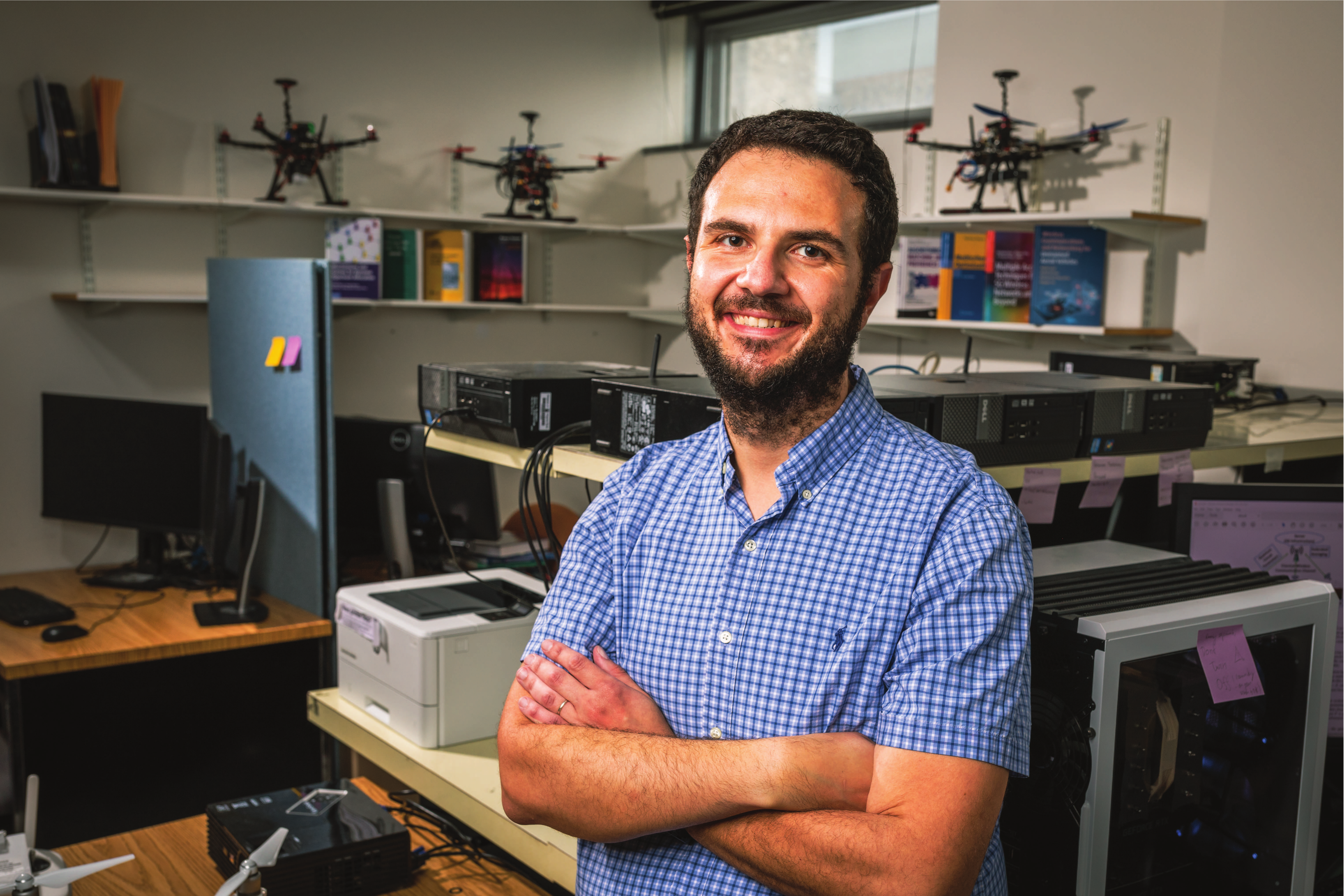}}]{Walid Saad} (S'07, M'10, SM'15, F'19) received his Ph.D degree from the University of Oslo
	in 2010. He is currently a Professor at the Department of Electrical and Computer
	Engineering at Virginia Tech, where he leads the Network sciEnce, Wireless, and Security
	(NEWS) laboratory. His research interests include wireless networks (5G/6G/beyond),
	machine learning, game theory, security, unmanned aerial vehicles, semantic
	communications) cyber-physical systems, and network science. Dr. Saad is a Fellow of the
	IEEE. He is also the recipient of the NSF CAREER award in 2013, the AFOSR summer
	faculty fellowship in 2014, and the Young Investigator Award from the Office of Naval
	Research (ONR) in 2015. He was the author/co-author of eleven conference best paper
	awards at WiOpt in 2009, ICIMP in 2010, IEEE WCNC in 2012, IEEE PIMRC in 2015, IEEE
	SmartGridComm in 2015, EuCNC in 2017, IEEE GLOBECOM in 2018, IFIP NTMS in
	2019, IEEE ICC in 2020 and 2022, and IEEE GLOBECOM in 2020. He is the recipient of the
	2015 and 2022 Fred W. Ellersick Prize from the IEEE Communications Society, of the 2017
	IEEE ComSoc Best Young Professional in Academia award, of the 2018 IEEE ComSoc
	Radio Communications Committee Early Achievement Award, and of the 2019 IEEE
	ComSoc Communication Theory Technical Committee. He was also a co-author of the 2019
	IEEE Communications Society Young Author Best Paper and of the 2021 IEEE
	Communications Society Young Author Best Paper. From 2015-2017, Dr. Saad was named
	the Stephen O. Lane Junior Faculty Fellow at Virginia Tech and, in 2017, he was named
	College of Engineering Faculty Fellow. He received the Dean's award for Research
	Excellence from Virginia Tech in 2019. He was also an IEEE Distinguished Lecturer in 2019-
	2020. He currently serves as an editor for the IEEE Transactions on Mobile Computing and
	the IEEE Transactions on Cognitive Communications and Networking. He is an Area Editor
	for the IEEE Transactions on Network Science and Engineering, an Associate Editor-in-Chief
	for the IEEE Journal on Selected Areas in Communications (JSAC) – Special issue on
	Machine Learning for Communication Networks, and an Editor-at-Large for the IEEE
	Transactions on Communications. He is the Editor-in-Chief for the IEEE Transactions on
	Machine Learning in Communications and Networking.\vspace{-1cm}
\end{IEEEbiography}

\begin{IEEEbiography}[{\includegraphics[width=1in,height=1.25in,clip,keepaspectratio]{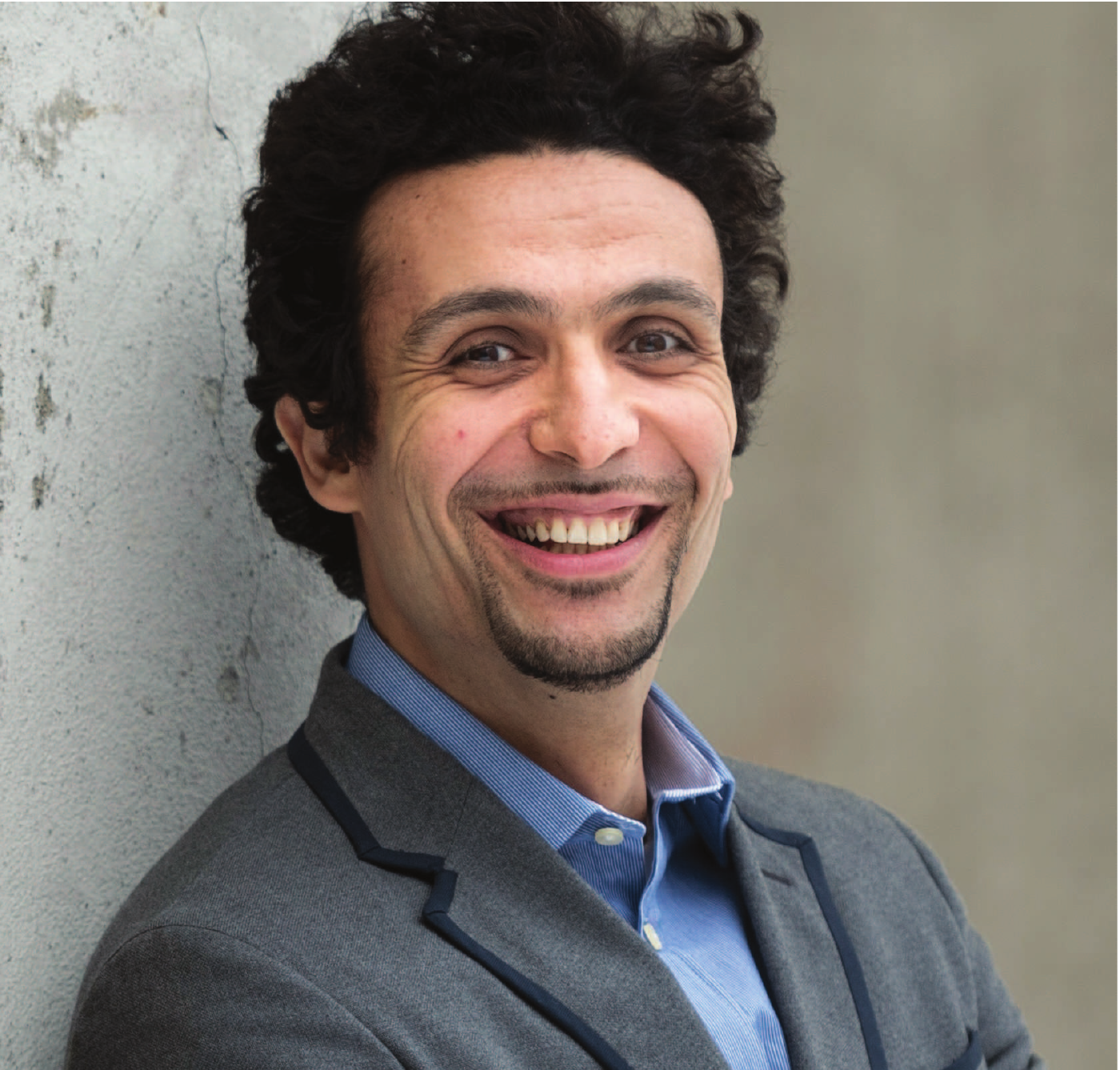}}] {Mehdi Bennis} (F'20) is a full (tenured) Professor at the Centre for Wireless Communications, University of Oulu, Finland and head of the intelligent connectivity and networks/systems group (ICON). His main research interests are in radio resource management, game theory and distributed AI in 5G/6G networks. He has published more than 200 research papers in international conferences, journals and book chapters. He has been the recipient of several prestigious awards including the 2015 Fred W. Ellersick Prize from the IEEE Communications Society, the 2016 Best Tutorial Prize from the IEEE Communications Society, the 2017 EURASIP Best paper Award for the Journal of Wireless Communications and Networks, the all-University of Oulu award for research, the 2019 IEEE ComSoc Radio Communications Committee Early Achievement Award and the 2020 Clarviate Highly Cited Researcher by the Web of Science. Dr Bennis is an editor of IEEE TCOM and Specialty Chief Editor for Data Science for Communications in the Frontiers in Communications and Networks journal. Dr Bennis is an IEEE Fellow. \vspace{-1cm}
\end{IEEEbiography}

\end{document}